\documentclass[twocolumn]{aastex63}

\usepackage[utf8]{inputenc} 
\usepackage[T1]{fontenc} 

\usepackage{amsmath}

\usepackage{bm} 

\usepackage{booktabs}

\let\tablenum\relax 
\usepackage{siunitx} 

\usepackage{placeins} 
\usepackage{float}

\submitjournal{ApJ}

\shorttitle{Requirements for gravitational collapse in planetesimal formation}
\shortauthors{Gerbig et al.}

\begin{document}


\title{Requirements for gravitational collapse in planetesimal formation --- the impact of scales set by Kelvin-Helmholtz and nonlinear streaming instability}

\correspondingauthor{Konstantin Gerbig} \email{gerbig@mpia.de}

\author{Konstantin Gerbig}
\affiliation{Max Planck Institute for Astronomy, Königstuhl 17, 69117 Heidelberg, Germany}
\affiliation{Department of Astronomy and Astrophysics, University of California, Santa Cruz, CA 95064, USA}

\author{Ruth A. Murray-Clay}
\affiliation{Department of Astronomy and Astrophysics, University of California, Santa Cruz, CA 95064, USA}

\author{Hubert Klahr}
\affiliation{Max Planck Institute for Astronomy, Königstuhl 17, 69117 Heidelberg, Germany}

\author{Hans Baehr}
\affiliation{Max Planck Institute for Astronomy, Königstuhl 17, 69117 Heidelberg, Germany}
\affiliation{Department of Physics and Astronomy, University of Nevada, Las Vegas, 4505 South Maryland Parkway, Las Vegas, NV 89154, USA}


\begin{abstract}

The formation of planetesimals is an unsolved problem in planet formation theory. A prominent scenario for overcoming dust growth barriers in dead zones is the gravitational collapse of locally over-dense regions, shown to robustly produce $\sim$100 km sized objects. Still, the conditions under which planetesimal formation occurs remain unclear. For collapse to proceed, the self-gravity of an overdensity must overcome stellar tidal disruption on large scales and turbulent diffusion on small scales. Here, we relate the scales of streaming and Kelvin-Helmholtz instability, which both regulate particle densities on the scales of gravitational collapse, directly to planetesimal formation. We support our analytic findings by performing 3D hydrodynamical simulations of streaming and Kelvin-Helmholtz instability and planetesimal formation.

We find that the vertical extent of the particle mid-plane layer and the radial width of streaming instability filaments are set by the same characteristic length scale, thus governing the strength of turbulent diffusion on the scales of planetesimal formation. We present and successfully test a collapse criterion: $0.1 Q \beta \epsilon^{-1}Z^{-1} \lesssim 1$ and show that even for Solar metallicities, planetesimals can form in dead zones of sufficiently massive disks. For a given gas Toomre-parameter $Q$, pressure gradient $\beta$, metallicity $Z$ and local particle enhancement $\epsilon$, the collapse criterion also provides a range of unstable scales, instituting a promising path for studying initial planetesimal mass distributions.  Streaming instability is not required for planetesimal collapse, but by increasing $\epsilon$, can evolve a system to instability. 

\end{abstract}



\section{Introduction} \label{sec:intro}

Protoplanetary disks around young stars are believed to mark the birthplace of planets. However, which disk processes govern planetesimal formation remains a central question. While \si{\micro \meter}-sized dust particles have been shown to efficiently grow \citep[e.g.,][]{Brauer2008}, further growth of \si{\meter}-sized pebbles is halted by fragmentation and rapid radial drift (e.g., \citealt{Birnstiel2012}; see \citealt{Chiang2010} for a review). Thus, processes to form planetesimals --- building blocks of planets defined as the smallest gravitationally bound objects --- must efficiently circumvent these growth barriers. While direct coagulation to planetesimals may be possible under certain conditions \citep[e.g.][]{Kataoka2013}, a substantial body of work has been developed suggesting that the ``m-size barrier" for growth is bypassed by the direct collapse to large planetesimals due to self-gravity. The simplest version of direct collapse--- gravitational fragmentation of the particle mid-plane \citep{Safronov1969, Goldreich1973} ---is thought to require super-solar metallicities \citep{Youdin_2002} due to vertical particle stirring induced by Kelvin-Helmholtz-instabilities \citep{Weidenschilling1980, Sekiya1998}. A promising solution to this problem is spontaneous gravitational collapse of locally over-dense particle clumps \citep{Johansen2006, Chiang2010, Simon2016}. Such clumps may be found in vortices \citep{Raettig2015, Manger2018} and zonal flows or other particle trap structures caused by instabilities in the gas disk \citep{Klahr2018, Pfeil2019} that convert a pebble flux to planetesimals \citep{Lenz2019, Gerbig2019}. 

In recent years, the streaming instability \citep{Youdin2005} has gained prominence as it was shown to be able to significantly increase local particle densities \citep{Johansen2007ApJ, Bai2010_ImplicationsPlanetesimals, Yang2014, Carrera2015, Yang2017, Schreiber2018} and thus bootstrap the formation of planetesimals with sizes of $\sim$ 100 \si{\kilo \meter} without the need for preexisting large-scale particle traps, as shown by e.g., \citet{Johansen2015, Simon2016, Schaefer2017, Nesvorny2019}. While \citet{Sekiya2018} recently showed that the level of particle concentration by streaming instability is regulated by just two parameters, the conditions required for collapse and planetesimal formation in the presence (or absence) of streaming instability remain unclear. 

To trigger local gravitational collapse, a particle clump must overcome both disruptive tidal shear and turbulent diffusion \citep{klahr_schreiber_2015, Klahr2019_criterion}. The former requires sufficiently high particle concentrations and the latter sufficiently low turbulent particle velocities, both of which are regulated by streaming and Kelvin-Helmholtz instability.

Kelvin-Helmholtz instabilities (KHI) in the particle stream are driven by the vertical gradient in orbital velocity intrinsic to disks with frictionally coupled dust and gas subject to a radial pressure gradient \citep{Weidenschilling1980, Nakagawa1986, Sekiya1998, Sekiya2000, Johansen2006KHI}. Kelvin-Helmholtz instability leads to vertical mixing and particle excitation and thus sets the vertical extent of the particle layer \citep{Weidenschilling1993, Youdin_2002, Chiang2008, Sekiya2018}. Streaming instability likewise is a drag instability \citep{Youdin2005, Jacquet2011, Umurhan2019}. Here, frictionally coupled particles exert a feedback force onto an epicyclic gas wave, which for the fastest growing wave modes is in resonance with the pressure-gradient-dependant dust streaming velocity \citep{Squire2018_RDIandSI, Zhuravlev2019}. The interplay of these two instabilities, albeit crucial for planetesimal formation, is not well studied numerically, in particular with self-gravity.

Thus, in this work, we employ the Pencil Code \citep{Brandenburg2002, Brandenburg2003} to conduct 3D fluid-dynamical simulations in the shearing-box approximation \citep{Goldreich1965, Balbus1992, Brandenburg1995} to numerically investigate the scales of streaming and Kelvin-Helmholtz instability. In contrast to e.g. \citet{Youdin_2007, Johansen2007ApJ, Schreiber2018}, we include vertical stellar gravity to capture particle settling and KHI-induced stirring. To verify analytic predictions for the vertical extent of the particle layer set by KHI \citet{Johansen2006KHI, Chiang2008}, we perform a brief parameter study, focusing on the effects of metallicity and of the gas disk's radial pressure gradient, which is the energy source of both instabilities. By then introducing self-gravity to our simulations, we test an updated version of the diffusion-limited collapse criterion in \citet{klahr_schreiber_2015, Klahr2019_criterion} and the conditions for planetesimal formation. This is in contrast to \citet{Shi2013}, who studied graviational collapse in the absence of a radial pressure gradient, i.e. without SI and KHI effects.  Finally, we put forth a more general collapse criterion that we derive by relating diffusion and local enhancement to the scales of streaming and Kelvin-Helmholtz instability.

The paper is structured as follows. In Sect.~\ref{sec:collapse_crit}, we review and expand upon the diffusion-limited collapse criterion from \citet{klahr_schreiber_2015, Klahr2019_criterion}. Sect.~\ref{sect:instabilitites_theo} reviews the effect of Kelvin-Helmholtz and streaming instability on particle densities in protoplanetary disk dead zones. In Sect.~\ref{sec:numerical_setup}, we discuss our numerical setup. Sects.~\ref{sect:vertical_scales} and \ref{sect:scales_si} present our numerical results for vertical and radial scales of particle overdensities in the absence and presence of streaming instability, respectively. In Sect.~\ref{sect:self_grav_and_pls_formation} we include self-gravity to numerically verify the collapse criterion. Finally, our conclusion, limitations and an outlook are presented in Sect.~\ref{sect:discussion}.
 

\section{Collapse criterion for local particle over-densities}\label{sec:collapse_crit}

The diffusion-limited collapse criterion for planetesimals from \citet{klahr_schreiber_2015} displays surprising similarities to the Toomre criterion for gas disk stability \citep{Toomre1964}. 
\citet{klahr_schreiber_2015, Klahr2019_criterion} present a collapse length scale appropriate for marginally unstable systems at the Hill density (see Eq.~\eqref{eq:density_larger_than_hill_density}).  We expand upon their result by recasting the interplay between diffusion, tidal shear, and self-gravity in a form familiar to the development of Toomre's $Q$ criterion for collapse.  In the process, we both account for systems that allow for multiple unstable modes and gain insight into the disk evolution processes that may lead up to planetesimal fragmentation, a subject that we return to in Sect.~\ref{sect:outlook_conclusion}.

We consider a particle cloud with density $\rho_\mathrm{c}$, mass $M_\mathrm{c}$ at distance from the star $a$, collapsing under its own gravity. A dust particle at distance $r$ from the cloud center is assumed to in-fall with its terminal velocity $v_\mathrm{term}$, i.e.,
\begin{align}
\label{eq:grav_vs_term_vel}
\frac{GM_\mathrm{c}}{r^2} = -\Omega\frac{v_\mathrm{term}}{\mathrm{St}}
\end{align}
where we introduced gravitational constant $G$, orbital frequency $\Omega$ and Stokes number $\mathrm{St}$, a dimensionless parameterization of the stopping time $t_\mathrm{stop}$ defined as
\begin{align}
\label{eq:def:stokesnumber}
    \mathrm{St} = t_\mathrm{stop}\Omega.
\end{align}
Here, $t_\mathrm{stop} = mw/F_\mathrm{D}$ for a particle of mass $m$ experiencing drag force $F_\mathrm{D}$ as it moves at velocity $w$ relative to the gas. For $\mathrm{St} < 1$, particles are coupled to the gas, whereas particles with $\mathrm{St}>1$ decouple from the gas. As such, the Stokes number quantifies the aerodynamic behavior as encoded in particle size and gas density \citep[see][or e.g., \cite{Chiang2010} for a review]{Epstein1924}.

Integration of Eq.~\eqref{eq:grav_vs_term_vel} yields a radius evolution and contraction timescale
\begin{align}
    t_\mathrm{con} = \frac{\Omega}{4\pi G \rho_\mathrm{c} \mathrm{St}},
\end{align}
on which the particle reaches the cloud center. 

\subsection{The diffusion limited collapse criterion} 

If the cloud is embedded in turbulent gas, and its particles are coupled to the gas i.e., $\mathrm{St} < 1$, particles will be subject to turbulent diffusion. We quantify the strength of this diffusion with the dimensionless parameter $\delta$, such that the diffusion coefficient is given by
\begin{align}
    D = \delta c_\mathrm{s}H.
\end{align}
Here we introduced sound speed $c_\mathrm{s}$ and gas pressure scale height 
\begin{align}
    H = \frac{c_\mathrm{s}
    }{\Omega}.
\end{align}
Diffusion is assumed to be isotropic. We discuss and test this assumption in  Appendix~\ref{sect:radial_vs_vertical_diff}. Following Fick's second law for diffusion, particles are diffused over the length scale $r$ with the diffusion timescale \citep{YoudinLithwick2007}
\begin{align}
\label{eq:diffusiontime}
    t_\mathrm{diff} = \frac{r^2}{\delta c_\mathrm{s}H}.
\end{align}
Thus, cloud collapse will be prevented by turbulent diffusion if $t_\mathrm{diff} < t_\mathrm{con}$, or
\begin{align}
\label{eq:timescalecomp}
    \frac{r^2}{\delta c_\mathrm{s}H} < \frac{\Omega}{4\pi G \rho_\mathrm{c} \mathrm{St}}.
\end{align}

Assuming the cloud extends to the vertical particle scale height $H_\mathrm{p}$, i.e., if $H_\mathrm{p} \approx r$, vertical integration of $\rho_\mathrm{c} $ yields the cloud particle column density $\Sigma_\mathrm{p,c}$, i.e.,
\begin{align}
\label{eq:cloud_density}
\rho_\mathrm{c} = \frac{\Sigma_\mathrm{p, c}}{2r}. 
\end{align}
Thus, we can transform Eq.~\eqref{eq:timescalecomp} to a diffusion-limited stability criterion
\begin{align}
\label{eq:diffusion_stability}
    r < \frac{\delta}{\mathrm{St}}\frac{c_\mathrm{s}^2}{2\pi G \Sigma_\mathrm{p, c}} =: L_\mathrm{diff}.
\end{align}
If the radius $r$ of a cloud with column density (mass per area) $\Sigma_\mathrm{p, c}$ falls below $L_\mathrm{diff}$, collapse will be prevented by turbulent diffusion.

\subsection{Cloud stability against tidal shear}

Whether a particle is gravitationally bound to the cloud or disrupted by tidal shear originating from the central star may be assessed by comparing the cloud's self-gravitational force to the force of tidal gravity.  Self-gravity dominates when 
\begin{equation}
\frac{GM_c}{r^2} > 3\frac{GM_*}{a^2}\frac{r}{a} \;\;,
\end{equation}
where $a$ is the separation between the cloud and the star.  This criterion may be conveniently re-written using the Hill-radius
\begin{align}
    r_\mathrm{H} = a \left(\frac{M_\mathrm{c}}{M_*}\right)^{1/3}.
\end{align}
The cloud will be stable against tidal shear if its density, $\rho_\mathrm{c}$, is larger then its local Hill density, i.e., \citep{klahr_schreiber_2015} 
\begin{align}
\label{eq:density_larger_than_hill_density}
    \rho_\mathrm{c} > \rho_\mathrm{H} := M_\mathrm{c}\left(\frac{4}{3}\pi r_\mathrm{H}^3\right)^{-1} = \frac{9}{4\pi}\frac{M_*}{a^3}.
\end{align}
Using Eq.~\eqref{eq:cloud_density}, the cloud radius $r$ for a given cloud column density $\Sigma_\mathrm{p,c}$ must not exceed
\begin{align}
\label{eq:hill_stability}
    r < \frac{2\pi}{9}\frac{G\Sigma_\mathrm{p,c}}{\Omega^2}=: L_\mathrm{hill}
\end{align}
for the cloud to be stable against tidal shear.

\subsection{Cloud radii subject to gravitational instability}

We combine the diffusion criterion in Eq.~\eqref{eq:diffusion_stability} and the tidal shear criterion in Eq.~\eqref{eq:hill_stability} to conclude that for local instability there must exist an $r$, such that $L_\mathrm{diff}< r < L_\mathrm{hill}$, i.e., 
\begin{align}
\label{eq:unstable_scales}
    \frac{\delta}{\mathrm{St}}\frac{c_\mathrm{s}^2}{2\pi G \Sigma_\mathrm{p, c}} < r < \frac{2\pi}{9}\frac{G\Sigma_\mathrm{p,c}}{\Omega^2}.
\end{align}
This is equivalent to 
\begin{align}
\label{eq:collapse_crit_noQorGamma}
    \frac{3}{2}\sqrt{\frac{\delta}{\mathrm{St}}} < \frac{\pi G \Sigma_\mathrm{p,c}}{c_\mathrm{s}\Omega}.
\end{align}
The right hand side of Eq.~\eqref{eq:collapse_crit_noQorGamma} is very reminisent of the inverse Toomre $Q$ parameter \citep{Toomre1964}
\begin{align}
\label{eq:def:Toomre_Q}
 Q := \frac{c_\mathrm{s}\Omega}{\pi G \Sigma_\mathrm{g}},
\end{align}
where $\Sigma_\mathrm{g}$ is the gas surface density. For $Q < 1$ the gas disk is subject to gravitational instability and fragmentation, since neither gas pressure nor tidal forces may prevent the gas disk's collapse \citep[see e.g.,][]{Baehr2017}.

Next, we can define the total metallicity
\begin{align}
\label{eq:def:disk_metallicity}
    Z := \frac{\Sigma_\mathrm{p}}{\Sigma_\mathrm{g}},
\end{align}
where $\Sigma_\mathrm{p}$ and $\Sigma_\mathrm{g}$ are spatially averaged particle and gas column densities respectively, i.e.
\begin{align}
    \Sigma_\mathrm{p} = \left\langle\int_{-\infty}^{\infty} \rho_\mathrm{p}\mathrm{d}z \right\rangle \text{\ and\ } \Sigma_\mathrm{g} = \left\langle\int_{-\infty}^{\infty} \rho_\mathrm{g}\mathrm{d}z\right\rangle,
\end{align}
where $\rho_{p}$ and $\rho_\mathrm{g}$ are particle and gas volume densities respectively. We express the local metallicity of the particle clump as
\begin{align}
\label{eq:local_metallicity}
    Z_\mathrm{c} = \frac{\Sigma_\mathrm{p,c}}{\Sigma_{g}} = \frac{\epsilon \Sigma_\mathrm{p}}{\Sigma_{g}} = \epsilon Z.
\end{align}
The dimensionless quantity $\epsilon$ characterizes potential radial and azimuthal enhancement in particle column density due to, for example, streaming instability. We note that for the purposes of our paper, $Z$ is defined by Eq.~\eqref{eq:def:disk_metallicity} at each location in the disk and thus does not necessarily correspond to the stellar [Fe/H] and could in fact vary temporarily and spatially due to dust drift and growth processes.

We can transform Eq.~\eqref{eq:collapse_crit_noQorGamma} to
\begin{align}
\label{eq:collapse_criterion_with_Q}
    1 > \frac{3}{2}\frac{Q}{\epsilon Z}\sqrt{\frac{\delta}{\mathrm{St}}} =: Q_\mathrm{p}.
\end{align}
Eq.~\eqref{eq:collapse_criterion_with_Q} is a criterion for instability. It neatly highlights the ambivalent effect of streaming instability on planetesimal formation, of both enhancing local particle densities \citep{Johansen_2007nature, Chiang2010, Simon2016} thus increasing $\epsilon$, while also providing the cloud with turbulent diffusion $\delta$ \citep{Schreiber2018}. Furthermore, high metallicities favor collapse, which is in line with e.g. \citet{Youdin_2002, Johansen_2009}. The critical value of $Q_\mathrm{p} = 1$, which is equivalent to $\rho_\mathrm{c} = \rho_\mathrm{H}$ and $L_\mathrm{hill} = L_\mathrm{diff}$, corresponds to the diffusion-limited collapse criterion of \citet{klahr_schreiber_2015, Klahr2019_criterion} and the attached marginally critical length scale, given by
\begin{align}
\label{eq:critical_lengthscale}
    r_\mathrm{crit} = \frac{1}{3}\sqrt{\frac{\delta}{\mathrm{St}}}H.
\end{align}
This can be derived by setting the three terms in Eq.~\eqref{eq:unstable_scales} equal to each other. However, for $Q_\mathrm{p} < 1$, more modes become unstable leading to a range of unstable scales that are both larger and smaller than the critical length scale. This may lead to the initial planetesimal size distributions seen in e.g. \citet{Simon2016, Schaefer2017, Simon2017, Abod2018, Nesvorny2019}. 

Metallicity $Z$ and dominating Stokes number $\mathrm{St}$ can be informed from dust-evolution models that contain growth, drift and fragmentation \citep[e.g.,][]{Birnstiel2010, Birnstiel2012, Lenz2019, Powell2019} and are therefore often treated as input parameters in planetesimal formation models, as is the Toomre $Q$, which parameterizes disk mass. On the contrary, the diffusion coefficient $\delta$ and metallicity enhancement $\epsilon$ are local properties which cannot be treated as input parameters. However, both can be related directly to the scales regulating local particle densities in planetesimal-forming disks.  These scales are set by drag instabilities, in particular the particle Kelvin-Helmholtz instability \citep{Sekiya1998, Johansen2006KHI, Chiang2008} and the nonlinear saturation of streaming instability \citep{Youdin2005, Squire2018_RDIandSI, Yang2014}. Thus, investigating how these two instabilities interplay and regulate local enhancement and diffusion is crucial for understanding planetesimal formation in the scenario of local gravitational collapse.


\section{Particle densities regulated by drag instabilities}
\label{sect:instabilitites_theo}

Protoplanetary disks contain gas and dust, both of which can be treated as fluids. The Euler equations of the two fluids are coupled via the drag force exerted by gas onto particles as well as its back-reaction by the particle flow onto gas. We define the position vector in the disk in cylindrical coordinates centered around the star, i.e. $\bm{r} = a \bm{\hat{r}} + \phi \bm{\hat{\phi}} + z \bm{\hat{z}}$. The set of hydrodynamic equations then reads \citep{Youdin2005}
\begin{align}
    \label{eq:gas_continuity}
    \frac{\partial \rho_\mathrm{g}}{\partial t} + \nabla \cdot \left( \rho_\mathrm{g}\bm{u}\right)  & = 0 \\
    \label{eq:gas_euler}
    \frac{\partial \bm{u}}{\partial t} + \left(\bm{u}\cdot\nabla\right)\bm{u} & =- \Omega^2 \bm{r}  + \mu \frac{\bm{w}}{t_\mathrm{s}}  - \frac{1}{\rho_\mathrm{g}}\nabla P\\
    \frac{\partial \rho_\mathrm{p}}{\partial t} + \nabla \cdot \left( \rho_\mathrm{p} \bm{v}\right) & = 0 \\
    \frac{\partial \bm{v}}{\partial t} + \left(\bm{v} \cdot \nabla \right)\bm{v} & = - \Omega^2 \bm{r} - \frac{\bm{w}}{t_\mathrm{s}}.
    \label{eq:particle_euler}
\end{align}
Here, we denote the gas and particle velocities as $\bm{u}$ and $\bm{v}$ respectively.  We have introduced gas pressure $P$, the local dust-to-gas ratio
\begin{align}
    \mu := \frac{\rho_\mathrm{p}}{\rho_\mathrm{g}},
\end{align}
and the relative velocity between particles and gas
\begin{align}
    \bm{w} := \bm{v} -\bm{u}.
\end{align}

While particles (in the absence of gas) orbit with Keplerian velocity
\begin{align}
\label{eq:keplerianvelocity}
v_\mathrm{K} = a\Omega,
\end{align}
radial stratification due to the pressure force in Eq.~\eqref{eq:gas_euler} causes the gas (in the absence of particles) to rotate with a sub-Keplerian velocity of
\begin{align}
    \label{eq:subkeplerianvelocity}
    v_\mathrm{K}\left(1 - \eta\right) = a\Omega \left(1 - \eta\right),
\end{align}
where the pressure support parameter, $\eta$, for a disk with aspect ratio 
\begin{align}
    h = \frac{H}{a} = \frac{c_\mathrm{s}}{v_\mathrm{K}}
\end{align}
is given by
\begin{align}
    \eta := - \frac{1}{2}h^2\frac{\mathrm{d}\ln \rho_\mathrm{g}}{\mathrm{d}\ln a} > 0.
\end{align}
Alternatively, the pressure gradient can be quantified by the dimensionless parameter
\begin{align}
    \label{eq:betadefinition}
    \beta := - h \frac{\mathrm{d}\ln \rho_\mathrm{g}}{\mathrm{d}\ln a} > 0,
\end{align} which leads to the relations
\begin{align}
    \eta = \frac{1}{2}h\beta,
\end{align}
and
\begin{align}
    \label{eq:etaatobetarelation}
    \eta a = \frac{1}{2}ha\beta = \frac{1}{2}\beta H.
\end{align}
Note that for a typical gas density profile exponent of around unity \citep[e.g.,][]{Andrews2009}, $\beta \sim h$. A fiducial value is $h = 0.1$, leading to $\beta = 0.1$ and $\eta = 0.05 \cdot h = 0.005$ \citep[e.g.,][]{Johansen2007ApJ}. We find $\beta$ to be the more convenient representation for the pressure gradient in our work, since length measurements become a fraction of gas scale height $H$ after Eq.~\eqref{eq:etaatobetarelation}, and are as such easily discernible. In the following sections, we discuss how the length scale $\frac{1}{2}\beta H$ in Eq.~\eqref{eq:etaatobetarelation} is characteristic for drag-instabilities, such as KHIs or SIs.

\subsection{Vertical particle extent regulated by Kelvin-Helmholtz instability}

\label{sect:KHI}

Eqns.~\eqref{eq:keplerianvelocity} and \eqref{eq:subkeplerianvelocity} can only describe particle and gas orbital velocities correctly as long as the two are not significantly subject to frictional coupling. For strongly coupled particles with $\mathrm{St} < 1$, the equilibrium azimuthal particle velocity $v_{0,y}$ of the combined fluid is \citep[e.g.][]{Nakagawa1986, Sekiya1998, Youdin_2002}
\begin{align}
    \label{eq:reducedparticlevelocity}
   v_{0,\phi}(a,z) = v_\mathrm{K}\left(1 - \frac{\eta}{1 + \mu(z)}\right).
\end{align}
We note that Eq.~\ref{eq:reducedparticlevelocity} depends on the local dust-to-gas ratio, $\mu$, such that the azimuthal velocity limits to Eq.~\ref{eq:subkeplerianvelocity} when gas dominates and to Eq.~\ref{eq:keplerianvelocity} when dust dominates. The restriction $\mathrm{St} < 1$ is valid in the context of the herein presented simulations and is also justified physically by dust evolution models like \citet{Birnstiel2010, Birnstiel2012}.

Since particles are not vertically supported by a pressure gradient, they settle towards the mid-plane, resulting in a strongly $z$-dependant dust-to-gas ratio $\mu(z)$. In the mid-plane, where $\mu(z)$ is maximal, Eq.~\eqref{eq:reducedparticlevelocity} implies that both the particles and the gas will orbit at velocities close to Keplerian, whereas in layers above the mid-plane their orbital velocities are reduced (by $\eta v_\mathrm{K}$ in dust-free layers) . The resulting vertical gradient in orbital velocity leads to Kelvin-Helmholtz instabilities (KHIs), sometimes known as particle vertical shearing instabilities, that vertically mix and stir up particles \citep{Sekiya1998, Sekiya2000, Sekiya2001, Youdin_2002, Chiang2008}. We will follow \cite{Chiang2008} in their analytical consideration of the KHI.

Whether or not the dust layer is subject to KHI is commonly assessed with the Richardson number $\mathrm{Ri}$, which compares buoyancy oscillations (for incompressible perturbations given by the Brunt-Väisälä frequency) to the rate of vertical shearing \citep{Chandrasekhar1961}. It is given by
\begin{align}
\label{eq:Ri_general}
    \mathrm{Ri} := \frac{\left({g_z}/{\rho}\right)\left(\partial \rho / \partial z\right)}{\left(\partial v_\phi/\partial z\right)^2},
\end{align}
where $g_z = -\Omega^2 z$ is the vertical component of stellar gravity, and $\rho = \rho_\mathrm{g} + \rho_\mathrm{p}$. If the Richardson number falls below a critical value, i.e. if
\begin{align}
    \mathrm{Ri} < \mathrm{Ri}_\mathrm{crit} = \frac{1}{4}
\end{align}
for Cartesian flows \citep{Chandrasekhar1961, Howard1973, Li2003, Chiang2008}, KHIs trigger and dust parcels are overturned and mixed. The exact value of the critical Richardson number is problem dependant. \citet{Johansen2006KHI} found a critical value of $\mathrm{Ri}_\mathrm{crit} \sim 1$ when numerically studying self-sustained Kelvin-Helmholtz instability in protoplanetary disks. 

As dust settles much closer to the mid-plane than gas $\partial \rho / \partial z \approx \partial \rho_\mathrm{p} / \partial z$. Hence, the Richardson number in protoplanetary disks becomes \citep{Chiang2008}
\begin{align}
\label{eq:Ri_ppd}
    \mathrm{Ri} = - \frac{\Omega^2 z}{\rho_\mathrm{p}+\rho_\mathrm{g}}\frac{ \left(\partial \rho_\mathrm{p} / \partial z\right)}{\left(\partial v_\phi/\partial z\right)^2} = - \frac{1}{\left(\eta a\right)^2} \frac{\left(1+\mu\right)^3}{\partial \mu / \partial z} z,
\end{align}
where we plugged in Eq.~\eqref{eq:reducedparticlevelocity}. For $\mathrm{Ri} = \mathrm{const}$, which is expected if settling times are much longer than KHI growth rates \citep{Johansen2006KHI}, integration of Eq.~\eqref{eq:Ri_ppd} yields a vertical profile for the dust-to-gas ratio $\mu(z)$ dependant on the mid-plane dust-to-gas ratio $\mu_0$, i.e. \citep{Chiang2008}
\begin{align}
    \mu(z) = \left[\frac{1}{\left(1+\mu_0\right)^2}+ \frac{1}{\mathrm{Ri}}\frac{z^2}{(\eta a)^2} \right]^{-1/2} - 1.
\end{align}

Solving $\mu(z) = 0$ for $z$ yields the maximum extent of the particle layer \citep{Chiang2008}
\begin{align}
    \label{eq:zmax}
    z_\mathrm{max} := \sqrt{\mathrm{Ri}} \frac{\sqrt{\mu_0^2 + 2 \mu_0}}{1+\mu_0}\eta a = \frac{\sqrt{\mathrm{Ri}}}{2}\frac{\sqrt{\mu_0^2 + 2 \mu_0}}{1+\mu_0}\beta H.
\end{align}
Eq.~\eqref{eq:zmax} predicts the vertical height of the particle layer for a given mid-plane dust-to-gas ratio and constant Richardson number. For $\mu_0 \gtrsim 1$, $z_\mathrm{max}$ limits to $0.5 \sqrt{\mathrm{Ri}}\beta H$.

Particles are expected to settle until the mid-plane is dense enough and the critical Richardson number is reached. Mixing and vertical excitation are now energetically favorable and counteract further settling, and thus particles are lifted to $z_\mathrm{max}$. When particles are stirred up, the Richardson-number of the flow rises to sub-critical values again such that the Kelvin-Helmholtz instability can not grow further. This leads to a self-regulating steady-state where the flow Richardson number returns to the critical value and the maximum extent is again given by Eq.~\eqref{eq:zmax}. As Eq.~\eqref{eq:zmax} assumes a constant $\mathrm{Ri}$-flow, we treat the steady-state flow Richardson number as a parameter and investigate which value best fits our numerical setup.

Note, that this consideration does not account for the high-density mid-plane cusp found by \citet{Sekiya1998, Youdin_2002, Gomez2005} for super-solar metallicities. Indeed, for sufficiently high metallicities, \citet{Youdin_2002} showed that this overdense particle mid-plane can by massive enough to gravitationally fragment and form planetesimals \citep[also see e.g.,][]{Chiang2010}. These enhancements in metallicity may be found in the inner disk due to radial drift of particles \citep{Youdin_2002}, in particle traps such as pressure bumps \citep[e.g.,][]{Rice2006, Pinilla2012, Taki2016} or as the disk dissipates due to photoevaporation \citep{Gorti2015}.

However, none of these processes are thought to generically operate and sufficiently increase metallicities across a broad range of disk environments.

\subsection{Streaming instability}

The streaming instability \citep{Youdin2005}, therefore, may offer a promising path of ubiquitously forming planetesimals \citep{Simon2016, Nesvorny2019}. This section briefly summarizes the nature of streaming instability, recent developments and its importance in the context of the collapse criterion presented in Eq.~\eqref{eq:collapse_criterion_with_Q}.

The streaming instability is a linear instability that occurs when the particles' back-reaction onto gas are taken into account \citep{Youdin2005, Jacquet2011, Zhuravlev2019}. For marginally coupled particles and local dust-to-gas ratios exceeding unity, the SI displays rapid growth rates, thus strongly enhancing local particle densities \citep{Carrera2015, Yang2017} and triggering planetesimal formation \citep{Bai2010_ImplicationsPlanetesimals, Johansen2015, Simon2016, Schaefer2017, Nesvorny2019}. Whether or not this critical dust-to-gas ratio can be achieved depends on the metallicity and pressure gradient \citep{Sekiya2018}, which determine the vertical particle height set by KHI. The functionality of the SI can be understood in the context of resonant drag instabilities, a class of instabilities where the relative dust-gas motion is in resonance with a pressure perturbation in the gas \citep{Squire2018RDI}. As showed by \citet{Squire2018_RDIandSI}, for the SI, the equilibrium solution for the relative dust-gas streaming velocity provided by \citet{Nakagawa1986} is in resonance with an epicyclic perturbation in the gas. 

Recently, the role of the linear SI in the context of planetesimal formation was cast into doubt by two additional considerations that were omitted in the stability analysis by \citet{Youdin2005} and likewise in most numerical work. First, the inclusion of external turbulence strongly suppressed SI growth rates, in particular on small scales \citep{Umurhan2019}. Secondly, \citet{Krapp2019} showed that the inclusion of multiple particle species also damps growth rates significantly. Still, in dead-zones, where turbulence is entirely self-induced by SI and KHI, and if the particle size distribution is sufficiently top-heavy and narrow as suggested by recent dust evolution models \citep[e.g.][]{Birnstiel2011, Okuzumi2012, Birnstiel2016}, SI remains an important processes and continues to be a key aspect of global models for planetesimal formation such as \citet{Drazkowska2016}. 

Therefore, understanding the ambivalent role of SI for gravitational collapse is crucial. On the one hand, its linear phase and the subsequent onset of turbophoresis\footnote{Turbophoresis \citep{Caporaloni1975} is the tendency of particles coupled to a fluid flow to accumulate in regions of low turbulence \citep[also see e.g.,][]{Reeks2014, Belan_2014}.} can concentrate particles and increase local densities to Hill-density and beyond. On the other hand, the self-induced turbulence intrinsic to the nonlinear saturation of SI also diffuses particles apart and thus limits collapse on small scales \citep{klahr_schreiber_2015, Klahr2019_criterion}. In fact, the fastest linearly growing modes are always on the smallest scales \citep{Youdin2005, Squire2018_RDIandSI, Umurhan2019} and therefore typically diffusion-limited. As the complex interplay of these oppositional processes during 3D nonlinear SI remains to be studied in detail, we propose a straight-forward estimate for a characteristic scale of SI-induced over-densities.

3D simulations of nonlinear SI universally feature large-scale azimuthally elongated filaments as linear modes are sheared apart \citep[see e.g.,][or Sect.~\ref{sect:scales_si}]{Johansen2012, Yang2014, Simon2016, Li2018, Abod2018}. Previously, their typical separation (feeding zone) has been investigated numerically by \citet{Yang2014}, and their radial extent has been connected to the operating scale of SI \citep{Umurhan2019}. In the following, we apply the Richardson criterion used to derive $z_\mathrm{max}$ for the KHI in Eq.~\eqref{eq:zmax} to SI filaments to derive an analogous estimate for a characteristic radial scale. 

To order of magnitude we may neglect Keplerian shear and the radial gradient in azimuthal velocity follows from a radially-dependent dust-to-gas ratio $\mu = \mu(x)$ that by construction is present in SI filaments. Motivated by the findings of \citet{Squire2018_RDIandSI}, we take the restoring force per unit mass as $g_x = -\Omega^2 x$ from radial epicyclic oscillations. Under the assumption of a constant Richardson number, we can exactly match the argument presented in Sect.~\ref{sect:KHI} and arrive at a maximum radial extent $x_\mathrm{max}$ for SI filaments given by
\begin{align}
\label{eq:Lmax}
    x_\mathrm{max} = z_\mathrm{max}(\mu_0 \geq 1) = \frac{\sqrt{\mathrm{Ri}}}{2}\beta H =: L_\mathrm{max}.
\end{align}
We point out, that while the simplification $\mu_0 \geq 1$ is per definition true as otherwise the SI would not have produced an over-dense filament in the first place, the assumption of a constant Richardson-number is likely not correct due to the asymmetry that arises when Keplerian shear is taken into account. On the scales of SI-filaments, the linearized Keplerian shear velocity is $\sim \eta v_\mathrm{K}$, comparable to the velocity difference due to the particle density gradient, and adds to the relative dust-gas streaming velocity on one side of the filament and subtracts on the other. Also, as peak densities during the nonlinear SI often significantly exceed unity, it seems likely that an analogous structure to the high-density mid-plane cusp found by \citet{Sekiya1998} forms, which is neither considered in the vertical nor the radial Richardson criterion. For these reasons, we stress that this radial scale has to be taken as a rough estimate.

Still, we show in Sect.~\ref{sect:self_grav_and_pls_formation} that having an analytic prediction for the radial scale, even if approximate, is very useful for retracing local diffusivity and enhancement to metallicity and pressure gradient, and thus for evaluating our collapse criterion in Eq.~\eqref{sec:collapse_crit} without a detailed knowledge of the turbophoretic state of the nonlinear SI.


\section{Numerical setup}\label{sec:numerical_setup}

\begin{deluxetable*}{ccccccccc}
	\tablecaption{List of performed simulations. $L_{x,y,z} = 0.4 H$ for all simulations. Note that the fiducial runs are listed multiple times. The star marks the quantity for which a parameter study is conducted. \label{tab:simulations}}
	
	\tablehead{ \colhead{Run(s)} & \colhead{$N_{x,y,z}$} &
	\colhead{$N_\mathrm{par}$} & \colhead{$H_\mathrm{p,0}[H]$} & \colhead{$\mathrm{St}$} & \colhead{$Z$} & \colhead{$\beta$} & \colhead{$Q$} & \colhead{Varied quantity}
    }
	
	\startdata
	Set\_128 & $128$ & 209715 & 0.025 & $0.005$ & * & $0.1$ & - & *: $Z \in \{0.0002, 0.002, 0.02 \}$ \\
	Set\_64 & $64$ & 262144 & 0.025 & 0.005 & * & $0.1$ & - & *:$Z \in \{0.0002, 0.002, 0.06, 0.02, 0.1, 0.2\}$\\
	FidRun\_64 & $64$ & 262144 & 0.1 & $0.2$ & $0.02$ & $0.1$ & - & -\\
	FidRun\_128 & $128$ & 209715 & 0.1 & $0.2$ & $0.02$ & $0.1$ & - & -\\
    Beta\_128 & $128$ & 209715 & 0.1 & $0.2$ & $0.02$ & * & - & *:$\beta \in \{0.05, 0.1\}$\\
	Z\_128 & $128$ & 209715 & 0.1 & $0.2$ & * & $0.1$ & - & *:$Z \in \{0.01, 0.02, 0.03\}$\\
	Beta\_64 & $64$ & 262144 & 0.1 & $0.2$ & $0.02$ & * & - & *:$\beta \in \{0, 0.05, 0.07, 0.1, 0.13, 0.17, 0.2\}$\\
	Z\_64 & $64$ & 262144 & 0.1 & $0.2$ & * & $0.1$ & - & *:$Z[10^{-3}] \in \{0.1, 0.2, 0.6, 1, 2, 8, 20, 200\}$\\
    Grav\_FidRun\_64 & $64$ & 262144 & 0.1 & $0.2$ & $0.02$ & $0.1$ & * & *:$Q \in \{31.9, 16.0, 8.0, 5.3, 4.0\}$ \\
    Grav\_FidRun\_128 & $128$ & 209715 & 0.1 & $0.2$ & $0.02$ & $0.1$ & * & *:$Q \in \{160.6, 31.9, 16.0, 8.0, 5.3, 4.0\}$ \\
    LowZ\_128 & $128$ & 209715 & 0.1 & $0.2$ & $0.01$ & $0.1$ & * & *:$Q \in \{2.0, 1.5, 1.1, 0.9, 0.8\}$ \\
    InitCond\_64 & $64$ & 262144 & * & $0.01$ & $0.02$ & $0.1$ & - & *:$H_\mathrm{p,0}[H]\in\{0.001, 0.01, 0.1, 0.2\}$\\
	\enddata
\end{deluxetable*}

For our numerical studies, we employ the Pencil Code\footnote{\url{http://pencil-code.nordita.org/}}, which uses sixth-order spatial derivatives and third-order Runge-Kutta integration in time to solve hydrodynamic equations on an Eulerian grid with Lagrangian particles  \citep[for details, also see ][]{Brandenburg2002, Brandenburg2003}. All presented simulations are conducted in a shearing-box approximation, i.e. a locally Cartesian coordiante frame that is co-rotating with $\Omega$ at distance $a$ to the central star, where $x,y \ll a$  \citep[for the derivation of shearing box approximation equations see e.g.,][]{Goldreich1965, Umurhan2004}. Here, the unperturbed azimuthal velocity in the local frame is given by 
\begin{align}
    u_{0,y} = - q \Omega x,
\end{align}
where
\begin{align}
    q = - \frac{\mathrm{d} \ln \Omega}{\mathrm{d}\ln a} = \frac{3}{2},
\end{align}
quantifies the linearized radial shear.

\subsection{Gas and particles}

For all simulations, we consider a non-magnetized gas obeying an isothermal equation of state
\begin{align}
    P = c_\mathrm{s}^2\rho_\mathrm{g}.
\end{align}
This is a good assumption for disk regions that are dominated by irradiation instead of accretion heating \citep{Kratter2011}. Initially, the gas is evenly distributed in the radial and azimuthal directions, while vertically stratified due to the vertical component of stellar gravity causing mid-plane sedimentation. 
We achieve sub-Keplerian gas velocities by artificially imposing a radial pressure gradient to support the gas. This is quantified by $\beta$ defined in Eq.~\eqref{eq:betadefinition}, which is added to the gas Euler equation (see Eq.~\eqref{eq:gas_euler_code}).  Boundary conditions are shear-periodic, periodic and periodic in $x,y$ and $z$ respectively. We defer to \citet{Li2018} for a discussion of the effect of vertical boundary conditions on streaming instability.

We seed the box with Lagrangian super-particles following a Gaussian with a scale height of $H_\mathrm{p,0}$. We test different values for $H_\mathrm{p,0}$ and discuss our choices for subsequent simulation runs in Appendix~\ref{sect:initial_scale_height_numerical_test}. Each super-particle is characterized by its position $\bm{x}_\mathrm{p} = (x_\mathrm{p},y_\mathrm{p}, z_\mathrm{p})$ and velocity $\bm{v} = (v_x,v_y, v_z)$ (measured with respect to $u_{0,y}$), and represents a swarm of identical physical solids interacting with the gas as a group. The Triangular Shaped Cloud (TSC) scheme \citep{hockney1988computer, Youdin_2007} is applied to smooth out super-particle properties to the neighboring grid cells. Particle properties are characterized by the input parameters of particle Stokes number $\mathrm{St}$ from Eq.~\eqref{eq:def:stokesnumber} and  disk metallicity $Z$ from Eq.~\eqref{eq:def:disk_metallicity}. Note that the Pencil Code up-scales particle masses such that the total dust-to-gas ratio in the boxes matches $Z$ even if the vertical domain size does not include the entire gas disk.

We consider particles with a fixed $\mathrm{St}$. For streaming instability for a mixture of different particle species, we defer to \citet{Schaffer2018}.

We also note that while the particles may display significant clumping in our simulations, the gas is comparatively static throughout all simulations.

\subsection{Evolution equations}

In our Pencil Code setup, the gas obeys the continuity equation
\begin{align}
    \frac{\partial \rho_\mathrm{g}}{\partial t} + \nabla \cdot \left(\rho_\mathrm{g} \bm{u}\right) + u_{0,y} \frac{\partial \rho_\mathrm{g}}{\partial y} = f_\mathrm{D}\left(\rho_\mathrm{g}\right),
\end{align}
where $f_\mathrm{D}\left(\rho_\mathrm{g}\right)$ represents an artificial hyper-diffusivity. The gas velocity $\bm{u}$ relatitive to $u_{0,y}$ is evolved via the equation of motion
\begin{align}
\label{eq:gas_euler_code}
\begin{split}
     & \frac{\partial \bm{u}}{\partial t}  + \left(\bm{u}\cdot\nabla\right)\bm{u} +  u_{0,y}\frac{\partial \bm{u}}{\partial y} = -c_\mathrm{s}^2 \nabla \ln \rho_\mathrm{g} + \Omega h \beta \bm{\hat{x}} \\ & +  \left(2\Omega u_y \bm{\hat{x}} - \frac{1}{2}\Omega u_x \bm{\hat{y}} - \Omega z \bm{\hat{z}}\right)  + \frac{\mu \bm{w}}{t_\mathrm{s}} + f_\nu \left(\bm{u},\rho_\mathrm{g}\right).
\end{split}
\end{align}
Here, the terms from left to right are: the velocity time derivative, advection due to velocity perturbation, advection due to shear flow, pressure gradient, artificially imposed centrifugal support due to global radial pressure gradient, the combination terms from linearized stellar gravity, centrifugal force, and Coriolis force, the back-reaction of the drag force exerted on gas by particles, and an artificial hyper-viscosity.

Hyper-viscosity and diffusivity are used to stabilize the code and smooth out steep gradients while preserving the power at the larger scales \citep[See Appendix B of][]{Yang2012}.

Similarly, the evolution equations for the particles read
\begin{align}
\frac{\partial \bm{x}}{\partial t} = - q\omega x_p \bm{\hat{y}} + \bm{v}
\end{align}
and
\begin{align}
\frac{\partial \bm{v}}{\partial t} = \left(2\Omega v_y \bm{\hat{x}} - \frac{1}{2}\Omega v_x \bm{\hat{y}} - \Omega z \bm{\hat{z}}\right) - \frac{\bm{w}}{t_\mathrm{s}}
\end{align}
with the key difference being\sout{,} that particles are not subject to pressure gradient.

\subsection{Self-gravity}

\begin{figure*}[htp]
\includegraphics[width = \textwidth]{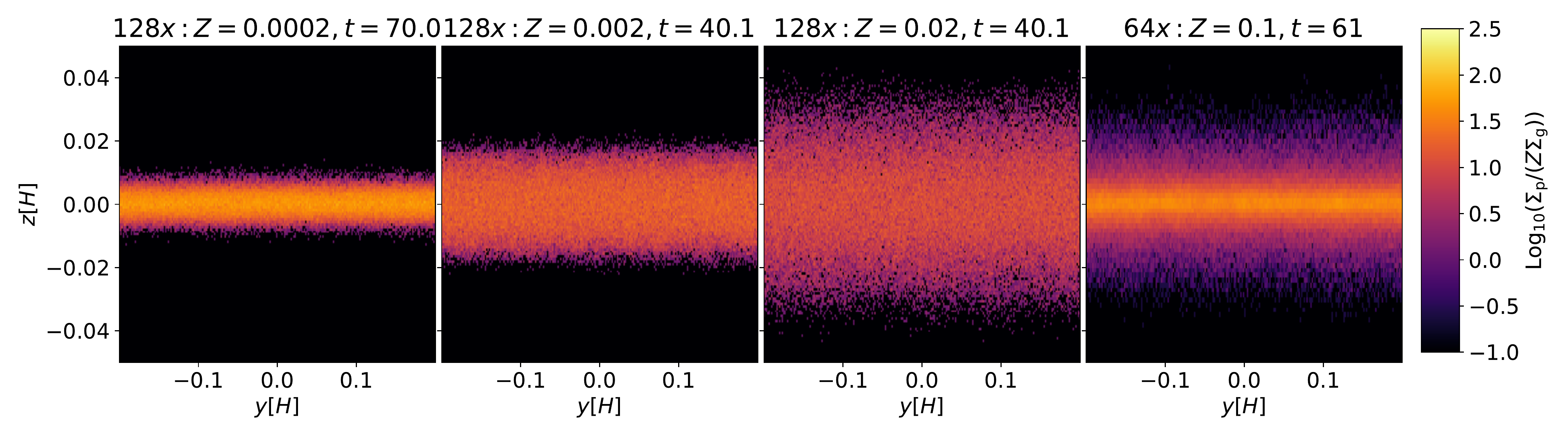}
\caption{Radially integrated column densities for selected settling simulations with $Z = 0.0002$ ($z_\mathrm{max} \approx 0.01 H$), $Z = 0.002$ ($z_\mathrm{max} \approx 0.02 H$), $Z = 0.02$ ($z_\mathrm{max} \approx 0.03 H$), and $Z = 0.1$ at the final snapshot of the simulation. For the first three simulations the final vertical extent of the particle layer well matches $z_\mathrm{max}$. In the $Z = 0.1$ simulation, a high-density mid-plane cusp develops and the vertical particle distribution is distinctly different to the simulations with lower metallicities. Also note that the $Z = 0.1$ simulation was only conducted with lower resolution in order to save computation time. For reference, $\frac{1}{2}\beta H = 0.05H$.  \label{fig:zmax_lowSt}}
\end{figure*}

\label{sect:numerical_selfgrav}

Self-gravity is implemented by solving the Poisson equation with the fast Fourier transform algorithm \citep{Gammie2001} containing gravitational softening. We consider the gravitational potential $\Phi(\bm{r})$ at arbitrary position $\bm{r}$ of both gas and particles. The relative strength of tidal shearing compared to self-gravity is parameterized with the Toomre $Q$ in Eq.~\eqref{eq:def:Toomre_Q} \citep{Toomre1964}.

This implies a value for the gravitational constant $G$ in code units, which is then plugged into the right hand side of the Poisson equation, i.e.
\begin{align}
    \Delta \Phi(\bm{r}) = 4\pi G \rho(\bm{r}) = \sqrt{\frac{8}{\pi}}\frac{\Omega^2}{Q \rho_\mathrm{g,0}}\rho(\bm{r}).
\end{align}
Note that $Q$ relates to the self-gravity parameter $\hat{G}$ used by \citet{Johansen2012, Simon2016, Schaefer2017} to parameterize self-gravity via
\begin{align}
    \hat{G} = \sqrt{\frac{8}{\pi}}\frac{1}{Q} \approx \frac{2}{Q}.
\end{align}
Lastly we note, that in our simulations even for $Q < 1$, the gas does not collapse as the chosen domain size does not include the critical unstable wave length of $2\pi H$.

\subsection{Units}

In our simulations we set the code units for angular frequency $\Omega = 1$, isothermal sound speed $c_\mathrm{s} = 1$, and initial mid-plane gas density $\rho_\mathrm{g, 0} = 1$. While the Pencil Code is agnostic to the choice of unit, we deem it useful to express our results in units that have physical meaning. Thus we choose the orbital period, which is related to the dynamical time via $2\pi\Omega^{-1}$, the gas scale height $H$, and the initial mid-plane gas density $\rho_{\mathrm{g},0}$ as time, length, and density units respectively. As the orbital frequency $\Omega$ is given by
\begin{align}
    \Omega = \sqrt{\frac{GM_*}{a^3}},
\end{align}
we can calculate scaling relations of our code units, i.e.
\begin{align}
    2\pi \Omega^{-1} & = \cdot \left(\frac{a}{1 \si{\ au}}\right)^{\frac{3}{2}}\left(\frac{M_*}{M_{\mathrm{sun}}}\right)^{-\frac{1}{2}} \si{yr} \\
    H & = 0.027\cdot\left(\frac{T}{180 \si{\ K}}\right)^{\frac{1}{2}}\left(\frac{a}{1 \si{\ au}}\right)^{\frac{3}{2}}\left(\frac{M_*}{M_{\mathrm{sun}}}\right)^{-\frac{1}{2}} \si{au},
\end{align}
where we used the isothermal sound speed given by
\begin{align}
    c_\mathrm{s} = \sqrt{\frac{k_\mathrm{B}T}{\Bar{m}}},
\end{align}
with Boltzmann constant $k_\mathrm{B}$ and a mean molecular weight of $\Bar{m} = 2.33 m_\mathrm{p}$, where $m_\mathrm{p}$ is the proton mass.

While the density scales freely with $\rho_\mathrm{g,0}$ for simulations where self-gravity is disabled, Eq. \eqref{eq:def:Toomre_Q} introduces a scaling relation once self-gravity is introduced. Thus, the density unit can be expressed via
\begin{align}
\rho_\mathrm{g,0} = 3 \cdot 10^{-7} \cdot \left(\frac{Q}{1}\right)^{-1} \left(\frac{a}{1 \si{\ au}}\right)^{-3}\left(\frac{M_*}{M_{\mathrm{sun}}}\right) \frac{\si{g}}{\si{cm}^3}.
\end{align}

\subsection{Simulation runs}

\label{sect:simulation_runs}

We conduct multiple simulations with various different parameter choices, displayed in Tab.~\ref{tab:simulations}. We follow \cite{Simon2016} and choose a domain size of $L_x = L_y = L_z = 0.4 H$, which allows us to sufficiently capture multiples of the characteristic scale $\beta H$. The majority of our simulations are based on a relatively small numerical grid with a size of  $64 \times 64 \times 64$, thus sacrificing resolution in favor of computational time. Additionally, we explore a higher resolution of $128 \times 128 \times 128$ for selected simulation setups. We denote the number of grid cells in radial, azimuthal and vertical direction with $N_x, N_y$ and $N_z$ respectively.

For these choices, the grid cell size is $\sim 0.006H$ and $\sim 0.003H$ respectively, which for $\beta = 0.1$ and typical viscosity values may not be enough to sufficiently resolve the fastest growing streaming instability wave length \citep{Umurhan2019}.  However, the marginally unstable scale for local gravitational collapse $r_\mathrm{crit}$ in Eq.~\eqref{eq:critical_lengthscale} \citep{Klahr2019_criterion} is for typical values just resolved (see Sect.~\ref{sect:diffusion_corr_time_scaleheight}). Nevertheless, the fact that our simulations, in particular those with low-resolution, are likely not fully converged is to be kept in mind when evaluating our numerical results. We discuss implications in Appendix.~\ref{sect:convergence}. For better resolved studies of pure streaming instability as well as gravitational  collapse, we refer to e.g., \citet{Yang2017, Schreiber2018, Sekiya2018} and \citet{Nesvorny2019, Klahr2019_criterion} respectively. 

For low resolution simulations, the grid is seeded with one super-particle per grid cell on average. High resolution simulations are seeded with $0.1$ super-particles per grid cell to save on computation time. As particles are expected to settle to a layer of thickness $\sim 0.04 H$ after Eq.~\eqref{eq:zmax}, corresponding to about a tenth of the domain size, the mid-plane layer will nevertheless have about one super-particle per grid cell.


\section{Vertical scales and diffusion coefficient}
\label{sect:vertical_scales}

\begin{figure}[htp]
\includegraphics[width = 0.47\textwidth]{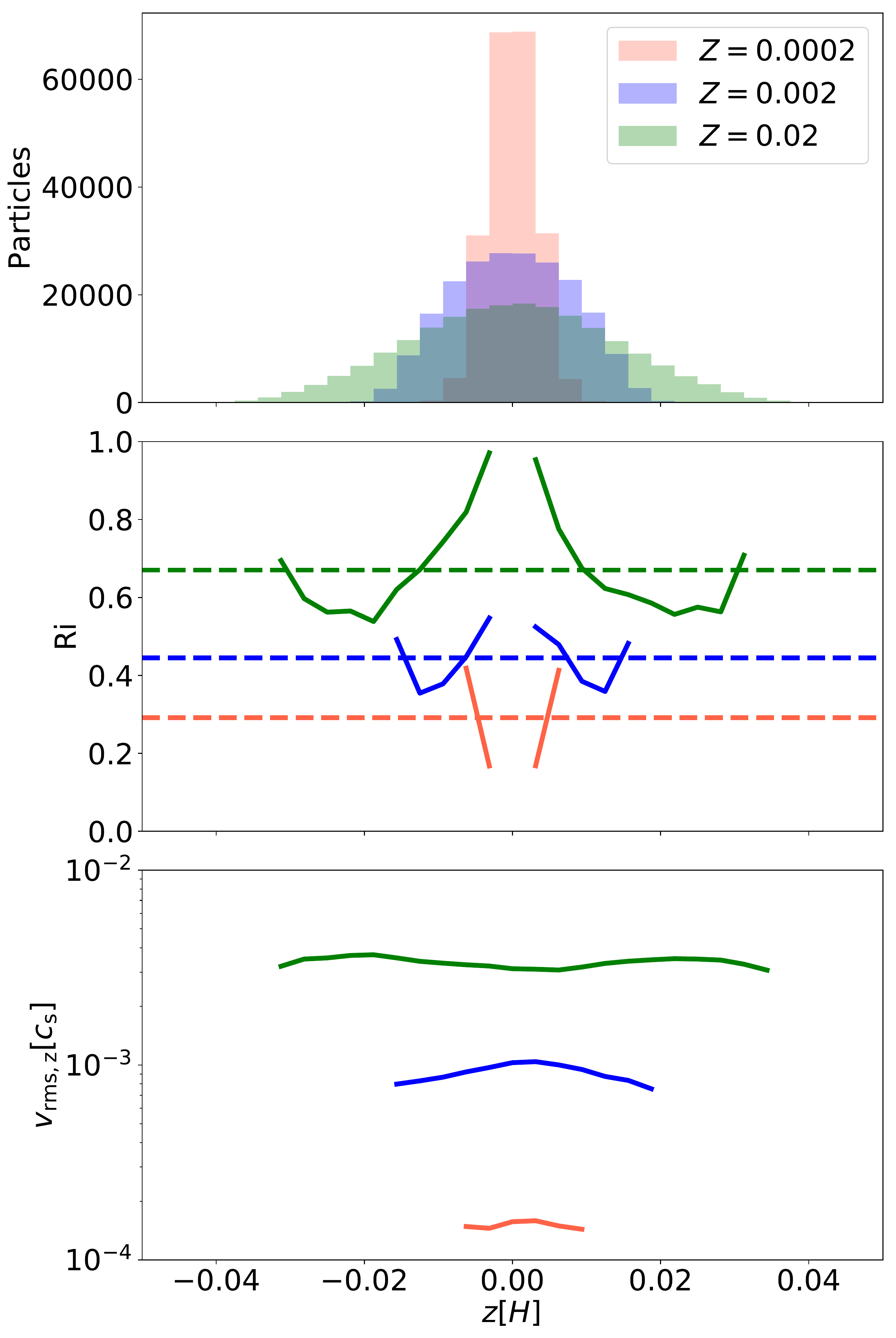}
\caption{Vertical profiles of particles from the simulations shown in the left three panels of Fig.~\ref{fig:zmax_lowSt}.  Vertical particle extent is well-correlated with the vertical particle velocity distribution. Above the midplane, the Richardson number is roughly constant with height in each disk.  Top panel: vertical particle distribution. Middle panel: flow Richardson number $\mathrm{Ri}$ after Eq.~\eqref{eq:Ri_from_sim} vs vertical coordinate $z$. Dashed lines indicate the mean Richardson number. Bottom panel: vertical component of root-mean-squared velocity after Eq.~\eqref{eq:vrms} vs $z$. All panels show high resolution settling simulations ($N_{x,y,z} = 128$, $\mathrm{St} = 0.005$) for three different metallicities $Z = 0.0002$, $Z = 0.002$, and $Z = 0.02$.\label{fig:Ri_vs_z}}
\end{figure}

Our first task is verifying that $z_\mathrm{max}$ from Eq.~\eqref{eq:zmax} indeed well describes the vertical extent of the particle layer. Further, we investigate how turbulent particle velocities relate to particle scale height $H_\mathrm{p}$, and therefore diffusion $\delta$ --- a key parameter in our collapse criterion in Eq.~\eqref{eq:collapse_criterion_with_Q} --- and $z_\mathrm{max}$.

For this purpose, we choose $\mathrm{St} = 0.005$ to slow down streaming instability growth rates \citep{Carrera2015, Yang2017}, and investigate the particle layer in the KHI-dominated case for four different values of metallicity. Particles are initialized with a scale height of $z_\mathrm{p,0} = 0.025 H$ and settle towards the mid-plane with settling time
\begin{align}
\label{eq:settling_time}
    t_\mathrm{set} = \frac{1}{\mathrm{St}\Omega}.
\end{align}
Appendix~\ref{sect:initial_scale_height_numerical_test} shows that the initial particle scale height does not affect the vertical extent in the self-regulated state.

We conducted a series of settling simulations with different metallicities (see Tab.~\ref{tab:simulations}), the final snapshot of four of which is displayed in Fig.\ref{fig:zmax_lowSt}. Note that for high metallicities of $Z \geq 0.1$, we only performed low resolution simulations to save computation time (see Appendix~\ref{sect:convergence} for a discussion of limitations due to the numerical resolution).

As expected from Eq.~\eqref{eq:zmax}, the vertical extent of the particle layer is $Z$-dependant. We identify three regimes. For low metallicities, the particles may settle very thin before entering the self-regulating regime ($z_\mathrm{max} \approx 0.01 H$, and $z_\mathrm{max} \approx 0.02 H$ for $Z = 2 \cdot 10^{-4}$ and $Z = 2 \cdot 10^{-3}$ respectively). Intermediate metallicities display the largest vertical extent ($z_\mathrm{max} \approx 0.03 H$ for $Z = 0.02$). For high metallicities the particle scale-height visibly falls below the analytic expectation from Eq.~\eqref{eq:zmax} ($z_\mathrm{max} \approx 0.025 H$ for $Z = 0.1$). There, we also recognize the high-density particle cusp around the mid-plane from \citet{Sekiya1998, Youdin_2002, Gomez2005}. While the Kelvin-Helmholtz instability attempts to stir up particles to $z_\mathrm{max}$, the mid-plane particle layer is so massive that stronger eddies are required for efficient particle lifting. Thus, only the surface layers can be stirred up effectively. As our consideration in Sect.~\ref{sect:KHI} does not well describe the vertical extent for this high-metallicity case, we will exclude this regime from further analysis in this section. 

\vspace{1cm}

\subsection{Flow Richardson number}

\begin{figure}[t]
    \includegraphics[width = 0.47\textwidth]{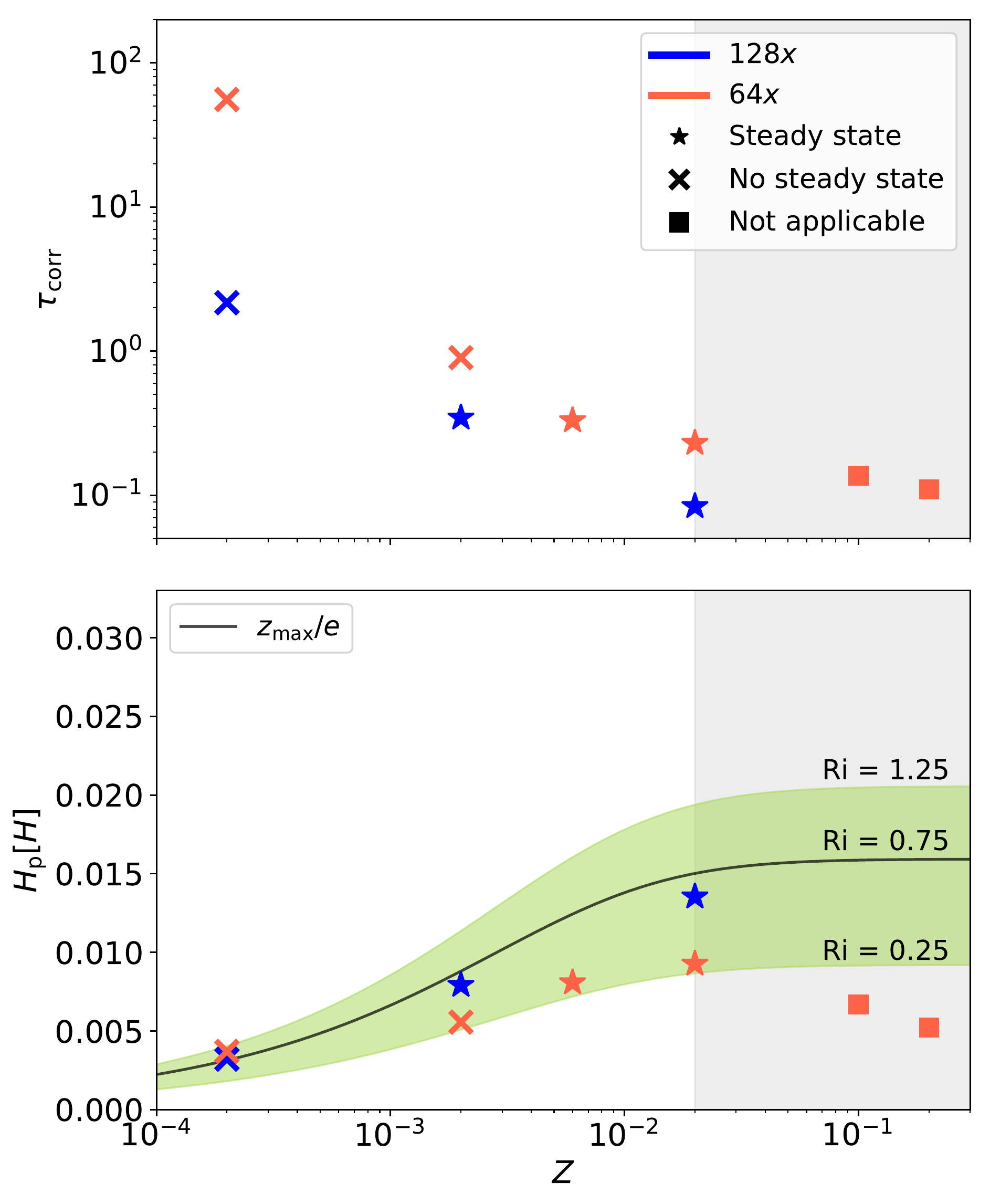}
    \caption{Correlation time after Eq.~\eqref{eq:correlation_time_from_vrms} (top panel) and particle scale height (bottom panel) during the final snapshot of settling simulations (top two rows of Tab.~\ref{tab:simulations}, $\mathrm{St} = 0.005$) are plotted against metallicity. Blue points correspond to simulations shown in Fig.~\ref{fig:Ri_vs_z}. The marker shape indicates if the system was able to reach a KHI-regulated steady state. Low metallicitiy simulations presumably have not fully settled yet and the correlation time is still decreasing. High metallicity simulations develop a high-density mid-plane cusp \citep{Sekiya1998, Gomez2005, Johansen2006KHI}, and the concept of a steady-state self-regulated by KHI is not applicable (approximately indicated by the gray region). The bottom panel depicts $z_\mathrm{max}/e$ from Eq.~\eqref{eq:zmax} for Richardson numbers of $\mathrm{Ri} = 0.75 \pm 0.5$.}
    \label{fig:tau_corr_and_Hp_vs_Z}
\end{figure}

\begin{figure*}[t]
\includegraphics[width = \textwidth]{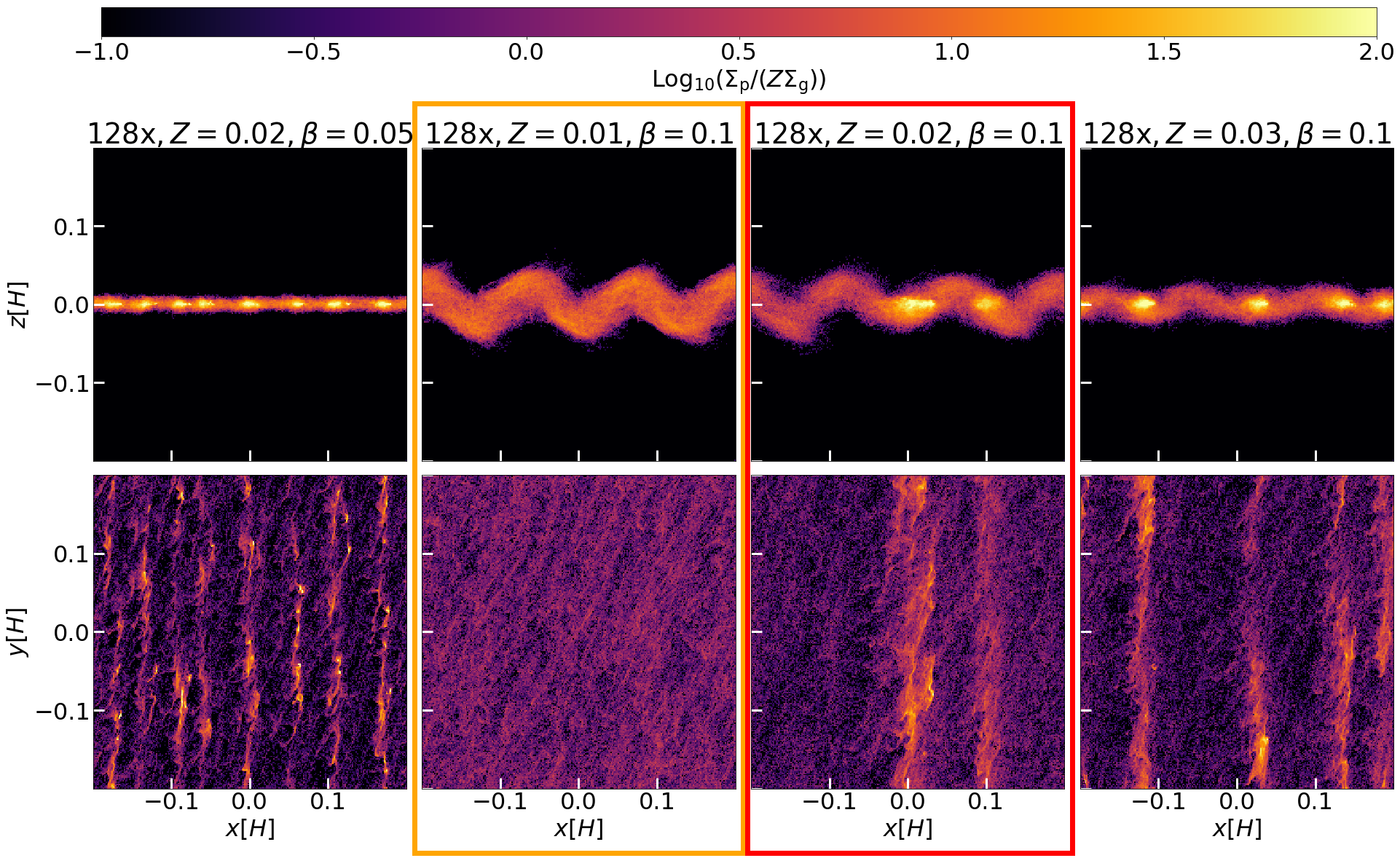}
\caption{Azimuthally integrated particle column densities (top panel) and vertically integrated particle column densities (bottom panel) for three high resolution simulations with $\mathrm{St} = 0.2$ at 40 orbits. The red rectangle marks the high resolution fiducial run, for which self-gravity was turned on at 60 orbits and the orange rectangle marks the high resolution run with $Z = 0.01$ for which self-gravity was turned on at 40 orbits (see Sect.~\ref{sect:self_grav_and_pls_formation}). \label{fig:scales_high_res}}
\end{figure*}

\begin{figure*}[t!]
\includegraphics[width = \textwidth]{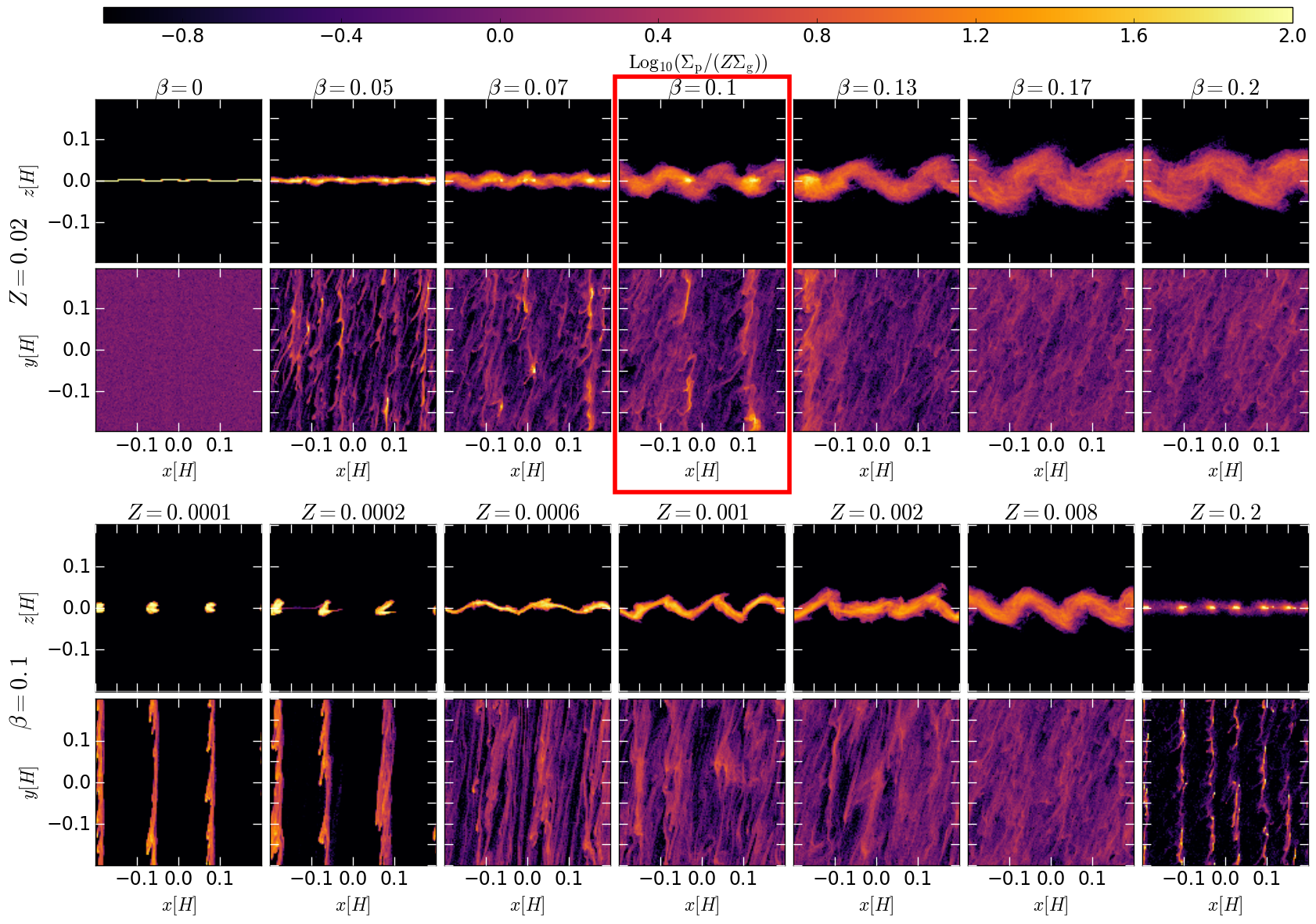}
\caption{Same as Fig.\ref{fig:scales_high_res} but for low resolution simulations. The red rectangle marks the low resolution fiducial run for which self-gravity was turned on. Note that this simulation would also fit between the two right most panels in the bottom row. \label{fig:scales_low_res}}
\end{figure*}

To appropriately relate particle scale heights to $z_\mathrm{max}$, it is worthwhile to first check whether the assumption of a constant Richardson number is appropriate for our system. Due to azimuthal and radial symmetry of the settling simulations, particle positions are projected onto the vertical axis. We calculate the Richardson number at height $z$ after Eq.~\eqref{eq:Ri_ppd} via
\begin{align}
\begin{split}
\label{eq:Ri_from_sim}
    \mathrm{Ri}(z) &= - \frac{4 z}{\beta^2 H^2} \frac{[1+\mu(z)]^3}{\partial \mu / \partial z}\\ &\approx - \frac{4 z }{\beta^2 H^2}[1+\mu(z)]^3 \frac{\Delta z}{\Delta \mu}.
\end{split}
\end{align}
This calculation is limited to our resolution. As such, $\Delta \mu$ is the difference in dust-to-gas ratio of two vertically adjacent slices, $\Delta z = L_z/N_z$, and $z$-positions of interest fulfill $z \in n \cdot L_z/N_z$ with $N_z > n \in \mathbb{N}$. To avoid statistical effects, we only calculate the Richardson number at a given $z$ if the two adjacent slices contain more particles than some threshold value, which was chosen to be $500$.

The height dependant Richardson number after Eq.~\eqref{eq:Ri_from_sim} for the final snapshot of high resolution settling simulations ($N_{x,y,z} = 128$, $\mathrm{St} = 0.005$) for three different metallicities $Z = 0.0002$, $Z = 0.002$, and $Z = 0.02$ is shown in the middle panel of Fig.~\ref{fig:Ri_vs_z}. Its top panel depicts the number of particles per slice in $z$. Note, that although the mid-plane contains about three time as many particles for $Z = 0.0002$ compared to $Z = 0.02$, the mid-plane dust-to-gas ratio is smaller as individual super-particle masses are lower by two orders of magnitude.

The vertical profile of the flow Richardson number qualitatively matches results by e.g., \citet{Johansen2006KHI, Bai2010_ImplicationsPlanetesimals}. Our estimate of the Richardson number supports findings by \citet{Johansen2006KHI} that the appropriate critical Richardson number is indeed greater than the literature value of $1/4$. Thus, while we continue to treat the Richardson-number as a parameter, we deem the assumption of it being constant required to derive $z_\mathrm{max}$ in Eq.~\eqref{eq:zmax} to be reasonable.

\vspace{1cm}

\subsection{Diffusion, correlation time and particle scale height}
\label{sect:diffusion_corr_time_scaleheight}

Next, we relate the vertical particle extent $z_\mathrm{max}$ to vertical diffusivity and show than root-mean-square velocities of particles behave accordingly. For a discussion of the radial diffusivity and our assumption of spherically-symmetric diffusion, see Appendix~\ref{sect:radial_vs_vertical_diff}.

The dimensionless diffusion coefficent $\delta$ is commonly measured using the root-mean-square velocity, the vertical-component of which is given by \citep{Carrera2015, Schreiber2018}
\begin{align}
\label{eq:vrms}
    v_\mathrm{rms, z} = \sqrt{\langle v_z^2\rangle} = \sqrt{\frac{1}{N_\mathrm{par}}\sum_i v_{z,i}^2},
\end{align} with $v_{z,i}$ being the vertical component of the velocity of particle $i$. It relates to the dimensionless diffusion coefficient via \citep{Johansen2006_diffusion, Schreiber2018}
\begin{align}
\label{eq:deltaz_from_vrms}
    \delta_\mathrm{z} = \tau_\mathrm{corr} \left(\frac{v_\mathrm{rms, z}}{c_\mathrm{s}}\right)^2,
\end{align}
where $\tau_\mathrm{corr}$ is the dimensionless correlation time. For $\mathrm{St} > 1$ particles, this correlation time is just given by the Stokes number $\mathrm{St}$ \citep{YoudinLithwick2007}. However, in our case for well-coupled particles with $\mathrm{St}< 1$, particle dispersion does not originate from random movement, but from gas turbulence. Thus, eddy turnover times will dictate correlation times \citep{YoudinLithwick2007}. For Kolmogorov turbulence, where particle turbulent kinetic energy is dominated by the largest eddies, whose turnover time is just the orbital time, then $\tau_\mathrm{corr} = 1$. In numerical simulations however, this is often not a good assumption. For example \citet{Schreiber2018} found values of $0.1 < \tau_\mathrm{corr} < 1$ depending on their simulation setup. Hence, we deem it worthwhile to measure appropriate correlation times for our simulation setup.

Equating settling time in Eq.~\eqref{eq:settling_time} and diffusion time in Eq.~\eqref{eq:diffusiontime} yields a well known expression for the particle scale height \citep{YoudinLithwick2007}
\begin{align}
\label{eq:particle_scale_height_from_deltaz}
    H_\mathrm{p} = \sqrt{\frac{\delta_z}{\mathrm{St}}} H.
\end{align}
In fact, Eq.~\eqref{eq:particle_scale_height_from_deltaz} is the definition of $\delta_z$ (and via our isotropy assumption also $\delta_x$) in the context of our work. Plugging in Eq.~\eqref{eq:deltaz_from_vrms} yields an expression for the dimensionless correlation time
\begin{align}
\label{eq:correlation_time_from_vrms}
    \tau_\mathrm{corr} = \mathrm{St} \cdot \left(\frac{H_\mathrm{p}/H}{v_\mathrm{rms,z}/c_\mathrm{s}}\right)^2.
\end{align}
We determine $v_\mathrm{rms}$ according to Eq.~\eqref{eq:vrms} and measure the particle scale height $H_\mathrm{p}$ using \citep{Carrera2015}
\begin{align}
\label{eq:measure_part_scaleheight}
    H_\mathrm{p} = \sqrt{\langle z^2\rangle} = \sqrt{\frac{1}{N_\mathrm{par}}\sum_i \left|z_i - \langle z \rangle \right|^2},
\end{align}
where $z_i$ is the vertical position of particle $i$, and $\langle z \rangle$ its mean over all $i$ (here $\langle z \rangle = 0$).

The dimensionless correlation times required to match root-mean-squared velocities of the three high-resolution settling simulations at the final snapshot to the respective particle scale heights are $\tau_\mathrm{corr}(Z = 0.0002) \approx 2$, $\tau_\mathrm{corr}(Z = 0.002) \approx 0.3$, $\tau_\mathrm{corr}(Z = 0.02) \approx 0.1$. This is in line with typical correlation times measured by e.g. \citet{Schreiber2018}. 

The top panel of Fig.~\ref{fig:tau_corr_and_Hp_vs_Z} plots the final snapshot correlation times vs metallicity for both low and high resolution simulations. The panel also distinguishes between simulations where the correlation time was able to reach a constant, non-evolving value within the simulation run time (steady state) and simulations where the correlation time was still evolving (no steady state), i.e. the particles are still in the settling phase and the flow has likely not yet reached the KHI-regulated equilibrium. The two simulations with super-solar metallicities of $Z = 0.1, Z = 0.2$ which develop the high-density mid-plane cusp are excluded from this classification.

Having verified that (vertical) root-mean-squared velocities correlate with the (vertical) diffusivity  and thus the particle scale height as anticipated, we proceed with Eq.~\eqref{eq:particle_scale_height_from_deltaz} and use  $H_\mathrm{p}$ to measure diffusivity. 

The bottom panel of Fig.~\ref{fig:tau_corr_and_Hp_vs_Z} plots the final snapshot particle scale-height vs metallicity $Z$. In order to overlay the analytic prediction from the self-regulated KHI steady state, we require a measurement for the mid-plane dust-to-gas ratio $\mu_0$. We find that the approximations $\mu_0 \approx (H / H_\mathrm{p})Z$ well matches our numerical result for $\mu_0$ \citep[also compare to Eq.~3 of][]{Sekiya2018}.

Moreover, to account for the fact that $z_\mathrm{max}$ by construction is the maximum particle extent, we must divide by an order unity factor to yield a quantity that relates to the scale height. The bottom panel of Fig.~\ref{fig:tau_corr_and_Hp_vs_Z} suggests, that $z_\mathrm{max}/e$ is a good match for low and intermediate metallicities that do not possess a high-density mid-plane cusp. 

Together with Eq.~\eqref{eq:particle_scale_height_from_deltaz}, $H_\mathrm{p} \approx z_\mathrm{max}/e$ implies
\begin{align}
\label{eq:delta_St_from_Lmax}
\sqrt{\frac{\delta_\mathrm{z}}{\mathrm{St}}} = \frac{H_\mathrm{p}}{H} \approx \frac{1}{e}\frac{z_\mathrm{max}}{H} = \frac{\sqrt{\mathrm{Ri}}}{2 e}\frac{\sqrt{\mu_0^2 + 2 \mu_0}}{1+\mu_0}\beta.
\end{align}
Note again, that the first equality in Eq.~\eqref{eq:delta_St_from_Lmax} is just the definition of $\delta_z$ in the context of our work. Relating $\delta_z$ to root-mean-squared velocities requires knowledge of correlation times, which we will defer to future work. The second equality relates the measured particle scale height to its analytic prediction from \citet{Chiang2008}, that was reviewed in Sect.~\ref{sect:KHI}. 

We can combine Eq.~\eqref{eq:delta_St_from_Lmax} with Eq.~\eqref{eq:critical_lengthscale} to find that, under the assumption of spherically symmetric diffusion (see Appendix~\ref{sect:radial_vs_vertical_diff}), the critical radius scale is just
\begin{align}
    r_\mathrm{crit} \approx \frac{1}{3}H_\mathrm{p} = \frac{\sqrt{\mathrm{Ri}}}{6 e}\frac{\sqrt{\mu_0^2 + 2 \mu_0}}{1+\mu_0}\beta H,
\end{align}
which for $\beta = 0.1$ and typical values for $\mu_0 \gtrsim 1$ and $\mathrm{Ri} \approx 1$ \citep[][or Fig.~\ref{fig:Ri_vs_z}]{Johansen2006KHI} is $r_\mathrm{crit} \approx  0.06 \beta H = 0.006 H$, and thus just resolved for the $N_{x,y,z} = 64$ simulations (also see Appendix~\ref{sect:convergence}).

\begin{figure}[t]
\includegraphics[width = 0.49\textwidth]{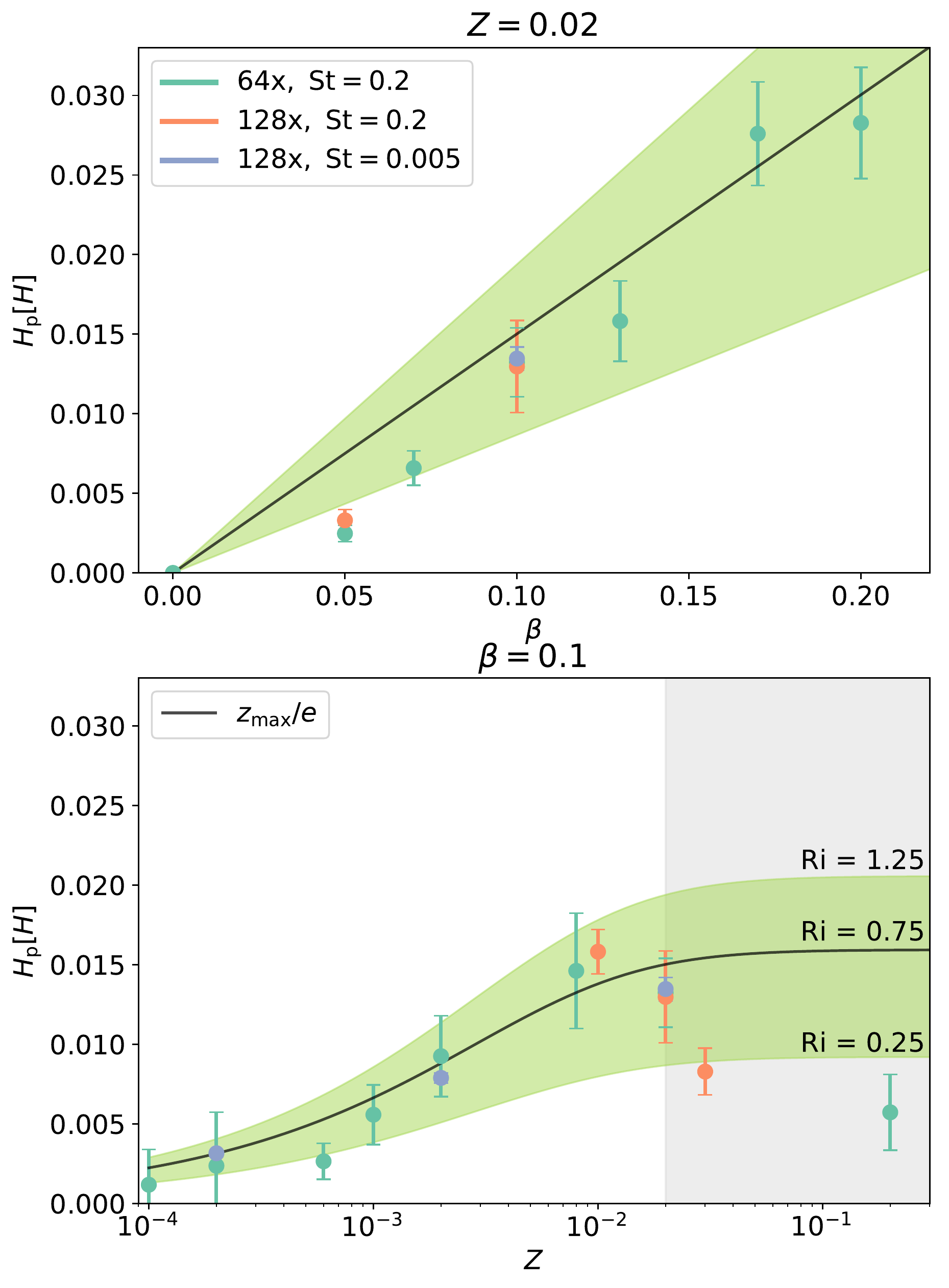}
\caption{Particle scale height vs $\beta$ (top panel) and $Z$ (bottom panel). Shown is the mean particle scale height at 40 orbits averaged over each $x$ for high and low resolution simulations depicted in Fig.\ref{fig:scales_high_res} (orange) and Fig.~\ref{fig:scales_low_res} (green) as well as high resolution settling simulations in Fig.~\ref{fig:zmax_lowSt} (blue). Error bars show the standard deviation originating from the radial average. Like in the bottom panel of Fig.~\ref{fig:tau_corr_and_Hp_vs_Z}, we plot the analytic expectation for $z_\mathrm{max}/e$ for different values of $\mathrm{Ri}$ for comparison. The gray region approximately indicates metallicities for which we do not expect the analytic expression to match our numerical results. \label{fig:vscale_vs_beta_and_Z}}
\end{figure}

\begin{figure}[t]
\includegraphics[width = 0.48\textwidth]{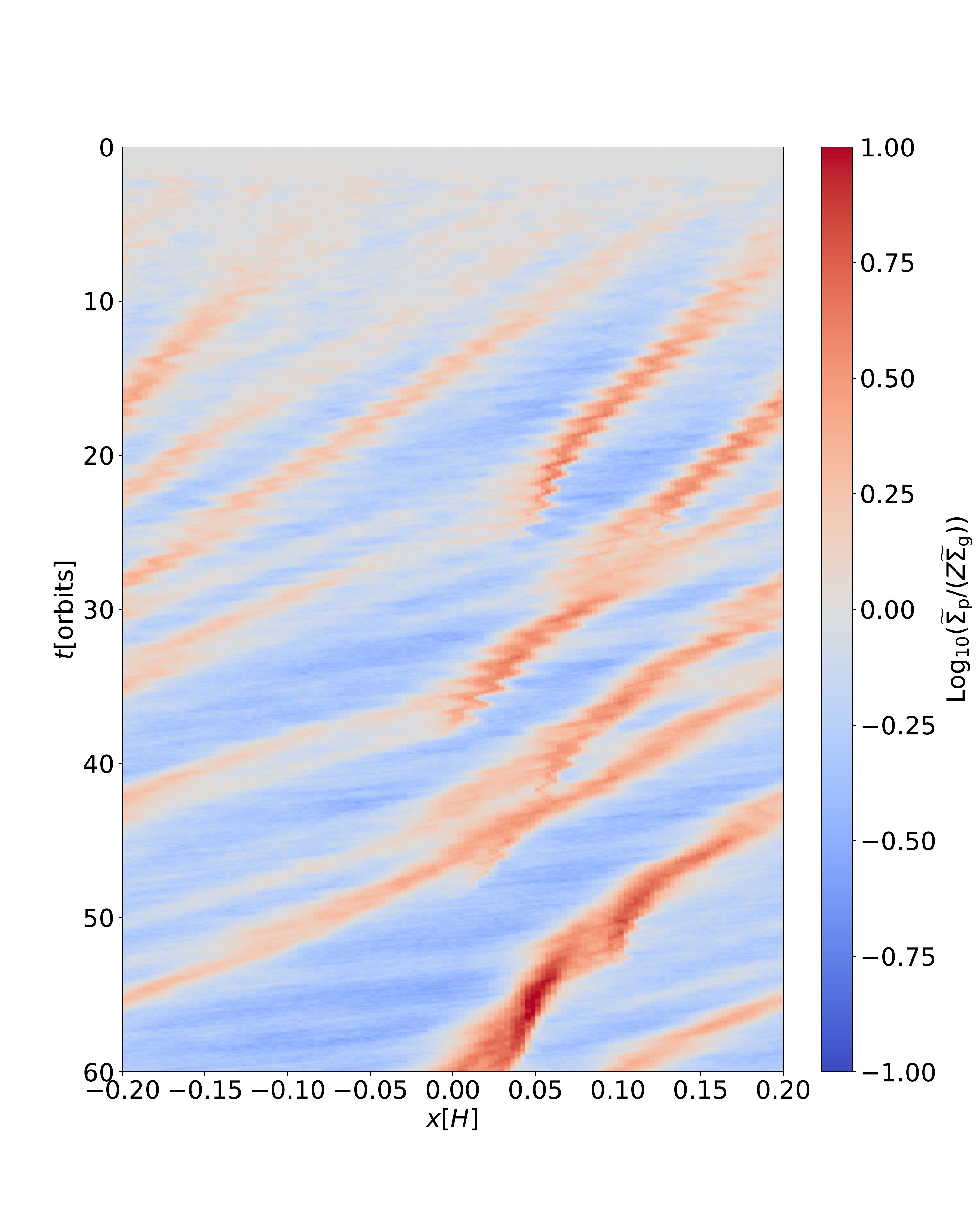}
\caption{Temporal evolution of slice density as defined in Eq.~\eqref{eq:sheet_density} for the high resolution fiducial run ($N_{x,y,z} = 128$, $Z = 0.02$). Note that red and blue regions are enhanced and depleted respectively with respect to the the unperturbed slice density. Inspired by similar figures in e.g.~\citet{Carrera2015, Li2018, Yang2018}. \label{fig:fil_tracks}}
\end{figure}


\section{Scales of streaming instability filaments}

\label{sect:scales_si}

Whereas the Kelvin-Helmholtz instability regulates vertical enhancements of the particle layer, the streaming instability does so in the radial direction by producing azimuthally elongated particle filaments \citep[see e.g.,][]{Simon2016, Sekiya2018}. In this section we investigate the scales of these filaments. First, we check how the presence of streaming instability alters the vertical extent of the particle layer. Then, we confirm our suspicion that the radial width of streaming instability filaments indeed never exceeds $L_\mathrm{max}$ from Eq.\eqref{eq:Lmax}.

All simulations presented in this section are performed with $\mathrm{St} = 0.2$, which roughly corresponds to the maximum particle size available in protoplanetary disks when considering drift and fragmentation limits \citep{Birnstiel2012, Powell2019}. 

Figures~\ref{fig:scales_high_res} and \ref{fig:scales_low_res} show the final snapshot of high resolution ($N_{x,y,z} = 128$) and low resolution ($N_{x,y,z} = 64$) simulations for a range of $\beta$ and $Z$ respectively.

\subsection{Vertical particle distribution in the presence of streaming instability}

The top panels of Figs.~\ref{fig:scales_high_res} and \ref{fig:scales_low_res} display azimuthally integrated particle densities. The wave-like structure characteristic to diagonal streaming instability wave modes emerges \citep[Compare to e.g.,][]{Li2018}. As this feature needs to be accounted for when calculating the particle scale height, $H_\mathrm{p,j}$ is calculated via Eq.~\eqref{eq:measure_part_scaleheight} at each slice $j$ in $x$ ($N_x$ slices with radial extent $L_x/N_x$) and then averaged, i.e.
\begin{align}
\label{eq:scaleheight_with_si}
    H_\mathrm{p} = \frac{1}{N_x}\sum_j H_\mathrm{p,j}.
\end{align}
The resulting final snapshot particle scale heights are depicted in Fig.~\eqref{fig:vscale_vs_beta_and_Z} against pressure gradient $\beta$ and $Z$. The high resolution settling simulations are plotted for comparison. The top panel of Fig.~\eqref{fig:vscale_vs_beta_and_Z} shows an approximately linear scaling of $H_\mathrm{p}$ in $\beta$, as expected from Eq.~\eqref{eq:zmax}. For the trivial case of $\beta = 0$, the relative dust-gas velocity is zero and both KHI and SI lose their energy source such that razor-thin settling occurs. The bottom panel of Fig.~\eqref{fig:vscale_vs_beta_and_Z} reproduces the bottom panel of Fig.~\ref{fig:tau_corr_and_Hp_vs_Z}. For low and intermediate metallicities $z_\mathrm{max}$ describes $H_\mathrm{p}$ well, even in presence of SI. For $Z > 0.01$, the high-density mid-plane cusp develops and the particle scale height decreases below the expected height. Thus, we expect the approximation in Eq.~\eqref{eq:delta_St_from_Lmax} to hold both in presence and in absence of streaming instability.

\subsection{Radial scales and enhancement}

\label{sect:radialscales}

The bottom panels of Figs~\ref{fig:scales_high_res} and \ref{fig:scales_low_res} display vertically integrated particle densities. Over-dense streaming instability filaments develop for $Z \geq 0.02$ and $0 < \beta \leq 0.13$, which is in line with e.g. \citet{Bai2010_pressuregradient}.

To quantify the width of a filament, we define the slice density as the vertically and azimuthally integrated density, i.e.,
\begin{align}
\label{eq:sheet_density}
\widetilde{\Sigma}_\mathrm{p}(x) := \int \Sigma_\mathrm{p}(x,y)\mathrm{d}y.
\end{align}
The temporal evolution of the radial distribution of the slice density for the high resolution fiducial run is shown in Fig.~\ref{fig:fil_tracks}. The slope of the filaments correlates with the local dust-to-gas ratio according to \citet{Nakagawa1986}. We also identify radial epicyclic oscillations in particular at early times. 

The radial extent of the a filament $\Delta x$ can be defined as
\begin{align}
\label{eq:radial_extent}
    \Delta x = x_+ - x_-,
\end{align}
with $x_- < x < x_+$ such that $\widetilde{\Sigma}_\mathrm{p}(x_-), \widetilde{\Sigma}_\mathrm{p}(x_+) < \widetilde{\Sigma}_\mathrm{p,0}<  \widetilde{\Sigma}_\mathrm{p}(x)$ holds true for all $x_- < x < x_+$. Here, $\widetilde{\Sigma}_\mathrm{p,0}$ is the mean particle slice density per grid cell extent $x_\mathrm{cell} = L_x / N_x$. Note that we must limit the radial resolution of this procedure to the gas grid cell size to remain self-consistent.

\begin{figure}[t]
\includegraphics[width = 0.5
\textwidth]{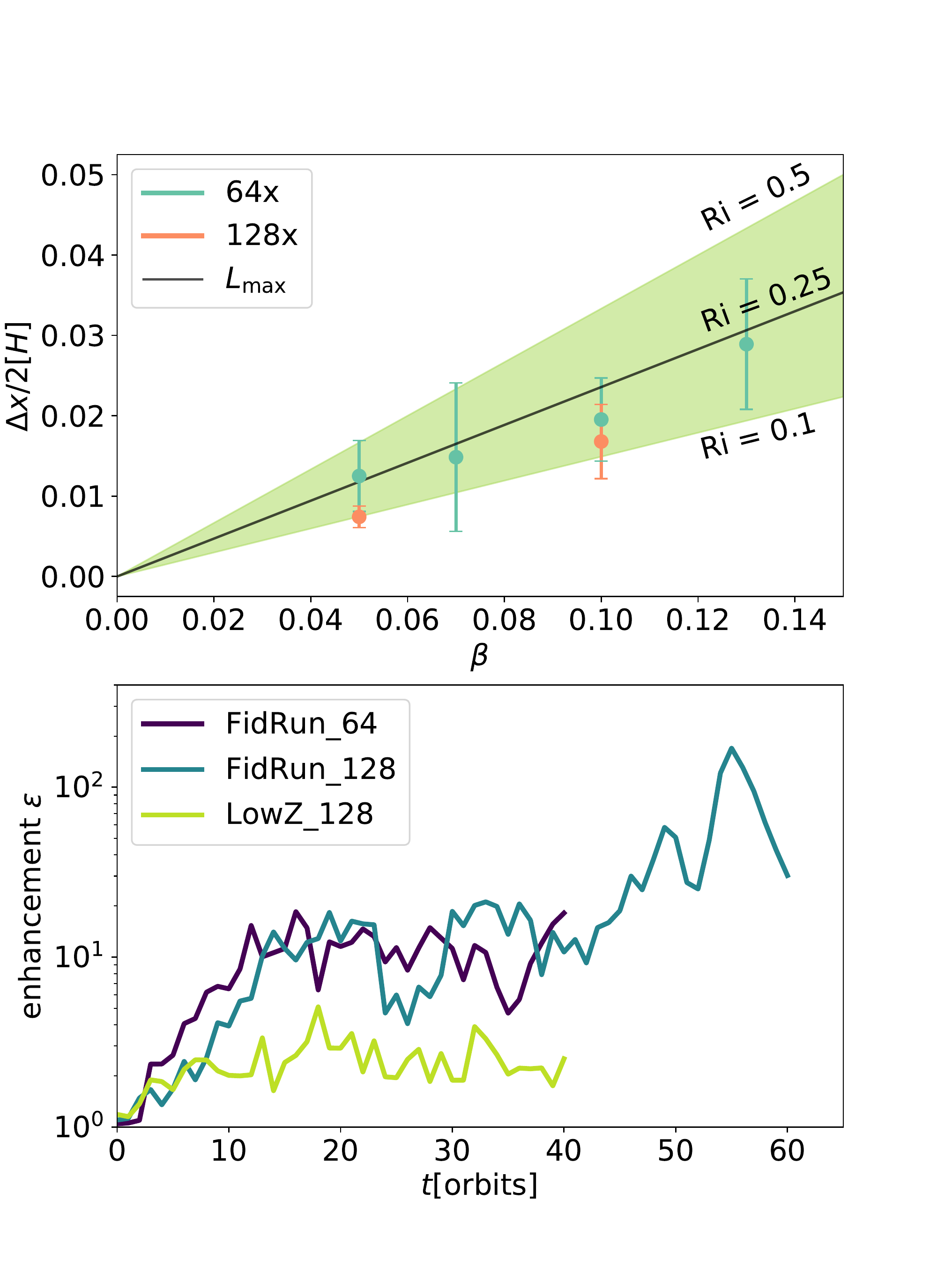}
\caption{Top panel: radial extent of the most dense filament after Eq.~\eqref{eq:radial_extent} vs $\beta$, overlayed with $L_\mathrm{max}$ from Eq.~\eqref{eq:Lmax} for three different Richardson numbers for simulations with $Z = 0.02$ that showed distinct SI-filaments (see main body for details). The extent $\Delta x/2$ was measured once per orbit and then averaged. The error bar is the resulting standard deviation  Bottom panel: temporal evolution of $\epsilon$ after Eq.~\eqref{eq:get_epsilon_from_sim} for the simulations  FidRun\_64 with $N_{x,y,z} = 64, Z = 0.02$, FidRun\_128 with $N_{x,y,z} = 128, Z = 0.02$, and LowZ\_128 with $N_{x,y,z} = 128, Z = 0.02$. We only plot times before self-gravity is turned on, which is at 40, 60, and 40 orbits respectively (see Sect.~\ref{sect:self_grav_and_pls_formation}).\label{fig:Delta_x_vs_beta}}
\end{figure}

The top panel of Fig.~\ref{fig:Delta_x_vs_beta} shows the radial extent of the most dense filament at the final snapshot vs pressure gradient $\beta$ for those simulations that developed distinct streaming instability filaments with $Z = 0.02$, i.e. FidRun\_128, Beta\_128 with $\beta = 0.05$, FidRun\_64, and Beta\_64 and $\beta \in \{0.05, 0.07, 0.1, 0.13\}$ (see also Tab.~\ref{tab:simulations}, and Figs.~\ref{fig:scales_high_res} and \ref{fig:scales_low_res}). The numerical data well match $L_\mathrm{max}$ in Eq.~\eqref{eq:Lmax}. To the extent that the vertical and radial scale heights of over-dense particle filaments are set by the same scale, Eq.~\eqref{eq:particle_scale_height_from_deltaz} implies that vertical and radial diffusivity must also be approximately equal and the approximation in Eq.~\eqref{eq:delta_St_from_Lmax} is valid for radial diffusivity as well.

The resulting dimensionless enhancement parameter $\epsilon$ defined in Eq.~\eqref{eq:local_metallicity} is estimated
\begin{align}
\label{eq:get_epsilon_from_sim}
    \epsilon = \frac{\Sigma_\mathrm{p,c}}{\Sigma_\mathrm{p}} \approx\left(\frac{1}{\Delta x \widetilde{\Sigma}_\mathrm{p,0}} \int_{x-}^{x_+}\widetilde{\Sigma}_\mathrm{p}(x)\mathrm{d}x\right)^p,
\end{align}
which by construction is larger than unity. Here, we introduced a correction coefficient $p\geq 1$. 

While the term in parenthesis assumes a azimuhthally isotropic filament, with radially uniformly distributed material, the correction exponent $p$ accounts for potential additional azimuthal enhancement, as well as the fact that a radially uniform distribution tends to underestimate the local enhancement. Note, that while the latter effect could be corrected via a multiplicative factor, the former requires an exponential coefficient. As it is challenging to account for these corrections rigorously, we treat them with a single correction exponent and set it to best match our simulation results. This procedure should be regarded as a first step and could be refined in future assessments of the collapse criterion.

The bottom panel of Fig.~\ref{fig:Delta_x_vs_beta} shows the evolution of $\epsilon$ for the three simulations, for which self-gravity is turned on in Section~\ref{sect:self_grav_and_pls_formation} at 40 orbits (FidRun\_64, $Z = 0.02$ and LowZ\_128, $Z = 0.01$) or 60 orbits (FidRun\_128, $Z = 0.02$). We set $p = 2.5$ for the two fiducial runs with $Z = 0.02$. Our methodology of measuring enhancement $\epsilon$ in Eq.~\eqref{eq:get_epsilon_from_sim} is specifically designed for enhancements within azimuthally elongated filaments. As such, in setups where those are not distinctly visible, as is the case for $Z = 0.01$, our procedure does not work, since the projection onto the $x$-axis removes all information about density enhancements that are rotated in the $x$-$y$-plane. Thus, we use $p = 4.5$ for the low metallicity run with  $Z = 0.01$, which best reproduces locally measured dust-to-gas ratios, to still get comparable values and assess the collapse criterion (see Sect.~\ref{sect:self_grav_and_pls_formation}). 

As expected for $Z = 0.01$, the local enhancement never exceeds an order unity factor of $\epsilon \sim 5$. This is in line with \citet{Carrera2015, Yang2017} who concluded that streaming instability can not effectively concentrate particles for this metallicity. For $Z = 0.02$, the enhancement parameter fluctuates around $10 \lesssim \epsilon \lesssim 100$. There is no qualitative difference between the two tested resolutions to be found. We attribute the peak in $\epsilon$ at 55 orbits for FidRun\_128 to statistical fluctuations typical to the nonlinear phase of streaming instability.


\section{Self-gravity and planetesimal formation}
\label{sect:self_grav_and_pls_formation}

\begin{figure}[htp]
\includegraphics[width = \linewidth]{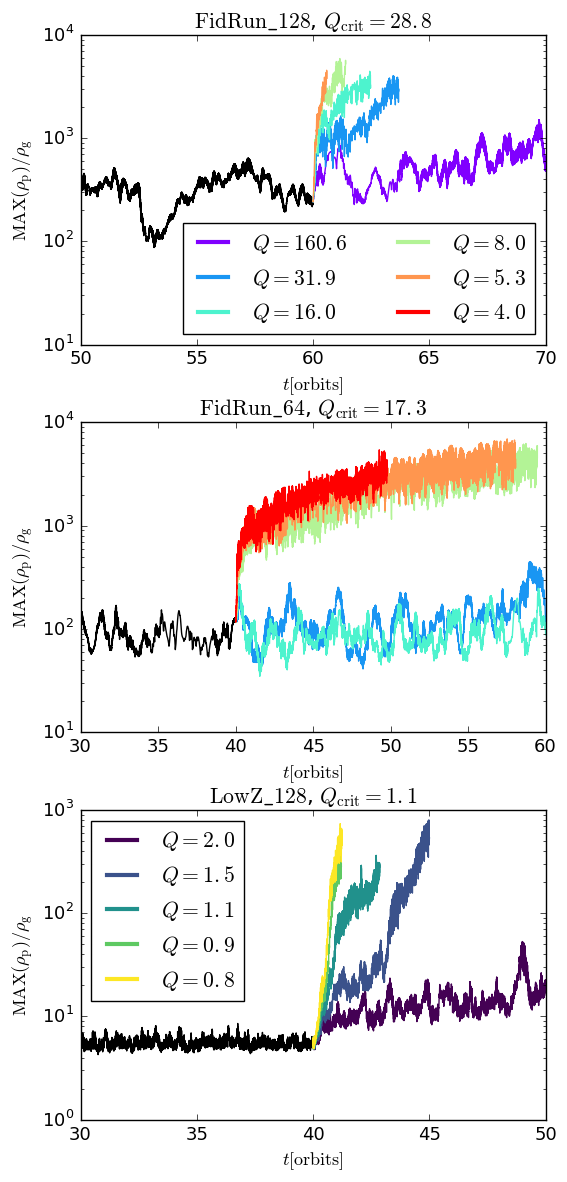}
\caption{The collapse criterion in Eq.~\eqref{eq:collapse_criterion_Gamma} (compare plots with $Q_\mathrm{crit}$ on top of each panel) predicts planetesimal formation. This figure shows the evolution of the maximum particle density for  simulations FidRun\_128 with $N_{x,y,z} = 128, Z = 0.02$ (top panel), FidRun\_64 with $N_{x,y,z} = 64, Z = 0.02$ (center panel), and LowZ\_128 with $N_{x,y,z} = 128, Z = 0.02$ (bottom panel) around the time self-gravity is turned on (at 60, 40 and 40 orbits respectively) for different values of the Toomre-parameter $Q$. \label{fig:Selfgrav_rhopmax}}
\end{figure}

\begin{figure*}[htp]
\includegraphics[width = \textwidth]{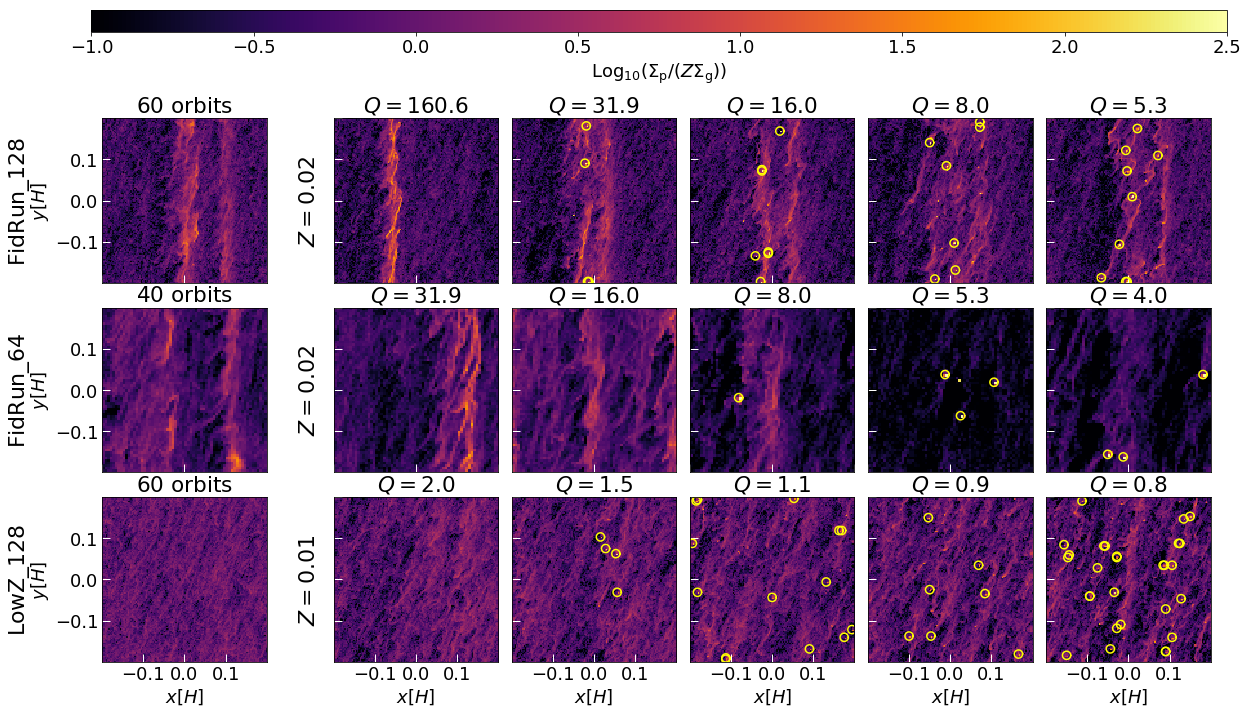}
\caption{Vertically integrated particle densities for self gravity simulations. Top, center and bottom row show simulations based on FidRun\_128 with $N_{x,y,z} = 128, Z = 0.02$, FidRun\_64 with $N_{x,y,z} = 64, Z = 0.02$, and LowZ\_128 with $N_{x,y,z} = 128, Z = 0.02$ respectively. The respective snapshot at the time self-gravity is turned on is depicted in the left column. Panels on the right five columns show the final snapshot for different $Q$-values (Compare with Fig.~\ref{fig:Selfgrav_rhopmax}). Yellow circles highlight planetesimal candidates found with a simple peak finder algorithm. For reference, the Toomre value of the Minimum Mass Solar Nebula (MMSN) is $\sim 30$ \citep{Weidenschilling1977MMSN, Hayashi1981}.\label{fig:planetesimals}}
\end{figure*}

Next, we numerically test the collapse criterion for planetesimal formation discussed in Sect.~\ref{sec:collapse_crit} and introduce self-gravity to our simulation. Equation~\eqref{eq:collapse_criterion_with_Q} defined $Q_\mathrm{p}$, a dimensionless quantity assessing the collapse of a particle cloud in analogy to Toomre-$Q$ for the gas disk. For collapse, $Q_\mathrm{p}$ must be less than one. Thus, after Eq.~\eqref{eq:collapse_criterion_with_Q}, for collapse, the Toomre-$Q$ of the system must fall below a critical value of $Q_\mathrm{crit}$ given by
\begin{align}
\label{eq:collapse_criterion_Gamma}
    Q < Q_\mathrm{crit} = \frac{2\epsilon Z}{3}\sqrt{\frac{\mathrm{St}}{\mathrm{\delta}}} = \frac{2\epsilon Z}{3} \frac{H}{H_\mathrm{p}},
\end{align} where we used Eq.~\eqref{eq:particle_scale_height_from_deltaz} and assumed $\delta = \delta_x = \delta_z$ (see Appendix~\ref{sect:radial_vs_vertical_diff}). Note that we choose to test for a critcial $Q$, rather than a critical $Q_\mathrm{p}$ as $Q$ is a direct input parameter in the Pencil-Code via the gravitational constant $G$ (see Sect.~\ref{sect:numerical_selfgrav}).

We verify the validity of this collapse criterion for three simulation setups: $N_{x,y,z} = 128, Z = 0.02$, $N_{x,y,z} = 128, Z = 0.01$, and $N_{x,y,z} = 64, Z = 0.02$ by testing 4 different values for $Q$ around a analytic estimation of $Q_\mathrm{crit}$. For all setups, $\beta = 0.1$ and $\mathrm{St} = 0.2$. We calculate $Q_\mathrm{crit}$ at the time self-gravity is activated by measuring $H_\mathrm{p}$ and $\epsilon$ via Eq.~\eqref{eq:scaleheight_with_si} (see Fig.~\ref{fig:vscale_vs_beta_and_Z}) and Eq.~\eqref{eq:get_epsilon_from_sim} (see bottom panel of Fig.~\ref{fig:Delta_x_vs_beta}) respectively. Similar to \citet{Simon2017, Schaefer2017, Nesvorny2019}, self-gravity is turned on at an arbitrary time during the non-linear saturation of streaming instability. This method must be regarded as a qualitative proof of concept and not as a quantitative verification of the collapse criterion. As such, we discuss potential problems with this procedure in Sect.~\ref{sect:outlook_conclusion}. 

Figure~\ref{fig:Selfgrav_rhopmax} shows the evolution of the maximum particle density for the three setups where self-gravity was included. The analytical prediction for $Q_\mathrm{crit}$ at the time self-gravity is turned on is noted in the top of each panel. Before self-gravity is turned on, $Q$ is not defined and all simulations behave identically.

For the high-resolution setups, the density increases significantly at the time self-gravity is turned on if $Q \lesssim Q_\mathrm{crit}$, which is indicative of collapse and planetesimal formation \citep[Compare to e.g.,][]{Simon2016, Klahr2019_criterion}. For the low resolution simulation, $Q = 16$ does not lead to collapse, which suggests that this simulation may not be converged (see Appendix~\ref{sect:convergence}). If the critical radius is not quite resolved, lower $Q$-values are required to render larger scales unstable.

The final snapshot of the vertically integrated particle density is shown in Fig.~\ref{fig:planetesimals}. There, planetesimal candidates are highlighted with yellow circles using a simple peak finder algorithm that identifes local maxima in the vertically-integrated particle surface density. We confirmed that planetesimal candidates indeed are gravitational bound by comparing their dispersive kinetic energy with their gravitational binding energy. We expect the initial planetesimal mass $M_\mathrm{pls}$ to scale with the cube of the size of the unstable region $r$, i.e. $M_\mathrm{pls} \propto \rho r^3$.  In general, the range of radii $r$ subject to instability will to depend on the value of $Q_\mathrm{p}$ in Eq.~\eqref{eq:collapse_criterion_with_Q}, but we expect the most unstable radius to be given by Eq.~\eqref{eq:critical_lengthscale}, which as we argue in this paper, is of order $\eta a$ --- the size of both the vertical particle height set by KHI and the radial width of SI filaments.  For a disk evolving gradually into an unstable configuration, we expect the density at initial collapse to be $\rho \approx \rho_\mathrm{H}$, so that $M_\mathrm{pls}$ scales as $\rho_\mathrm{H} (\eta a)^3$ \citep[also compare to][]{Chiang2014}. A detailed numerical investigation of the initial mass function in the context of the herein presented collapse criterion would require more sophisticated clump finder \citep[see e.g.][]{Li2019} and is subject to future work.

Fig.~\ref{fig:planetesimals} shows, that collapse can occur also in the absence of streaming-instability-induced clumping for a metallicity of $Z = 0.01$, if $Q$ is super-critical \citep[this is in contrast to e.g.,][]{Johansen_2009}. Likewise, even if streaming instability produces high local particle densities (for $Z = 0.02$), collapse does not occur if $Q$ is too high. Due to $Q \propto \rho_{g,0}^{-1}$, this implies that massive disks can form planetesimals in dead-zones without the prerequisite of streaming instability and that very low-mass disks cannot form planetesimals even if they are metal-rich.


\section{Discussion}
\label{sect:discussion}

\subsection{Towards an universal criterion for planetesimal formation}

We numerically showed, albeit only for a limited range of parameters, that  local gravitational collapse to form a planetesimal requires the diffusion- and tidal-shear-limited collapse criterion presented in Eq.~\eqref{eq:collapse_criterion_with_Q} to hold true. Applying the criterion to global disk models is not obvious as both $\epsilon$ and $\delta$ are local properties that are potentially dependent on resolution and included physics. However, in Sect.~\ref{sect:vertical_scales} we showed that the ratio $\delta/\mathrm{St}$ can be related to $L_\mathrm{max}$ via Eq.~\eqref{eq:delta_St_from_Lmax}. Thus, the collapse criterion in Eq.~\eqref{eq:collapse_criterion_with_Q} can be expressed as
\begin{align}
\label{eq:collapse_in_dead_zones}
    1 \gtrsim \frac{3}{2}\frac{Q}{\epsilon Z}\frac{\sqrt{\mathrm{Ri}}}{2e}\frac{\sqrt{\mu_0^2 +2\mu_0}}{1+\mu_0}\beta \approx \frac{1}{10}\frac{Q}{\epsilon}\frac{\beta}{Z} = Q_\mathrm{p}.
\end{align}
If $Q_\mathrm{p} \lesssim 1$, i.e. for small pressure gradients and $Q$-values, and for large metallicities and local enhancements, we expect collapse and planetesimal formation. This Toomre-like formulation of the collapse criterion highlights that cloud collapse and planetesimal formation do not fundamentally
depend on the presence of streaming instability and interpreting streaming instability as a required catalyst for planetesimal formation is not appropriate. Instead, collapse is assessed by comparing disk mass (quantified by $Q$), local dust content $\epsilon Z$ and pressure gradient $\beta$. In fact, \citet{Sekiya2018} recently showed that combining pressure gradient and metallicity to a single parameter appropriately describes streaming instability clumping, so it is no surprise that this ratio also appears in our collapse criterion.

The criterion in Eq.~\eqref{eq:collapse_in_dead_zones} can also be utilized to assess if it is easier to collapse the gas disk or local particle clumps. If 
\begin{align}
    \frac{Q_\mathrm{p}}{Q} \approx \frac{1}{10}\frac{\beta}{\epsilon Z} \lesssim 1,
\end{align}
particle collapse is easier. For our fiducial pressure gradient value of $\beta = 0.1$, and for no local enhancements ($\epsilon = 1$), this would be the case for $Z \gtrsim 0.01$.

\subsection{Limitations and other procceses potentially affecting the particle layer}

\label{sect:other_instabilities}

By replacing diffusivity with particle scale height, one also eliminates the explicit dependency of the collapse criterion on Stokes number. We point out, that Stokes number still implicitly effects the collapse criterion by influencing the local enhancement $\epsilon$ due to streaming instability. Our work is most appropriate for scenarios of narrow, top-heavy particle size distribution. The validity of the collapse criterion for different Stokes numbers as well as for a particle size distribution \citep[compare to e.g.,][]{Schaffer2018} which may strongly damp SI growth rates \citep{Krapp2019} remains to be studied. Similarly, we defer a high-resolution test to future work (also see Appendix~\ref{sect:convergence}).

Although the physical origin of KHI investigated in \cite{Chiang2008} and in our work is vertical shear, we explicitly abstain from referencing to it as vertical shear instability. This term is commonly used for a gas-only instability active in protoplanetary disks that are able to cool sufficiently fast to allow the development of vertical gas perturbations \citep[see e.g.][]{Urpin1998, Nelson2013, Lin2015, Pfeil2019}. Here, the vertical gradient in orbital frequency is not caused by dust-gas coupling, but instead by radial variations in temperature and entropy \citep{Barker2015}, leading to vertically elongated modes \citep{Arlt2004, Klahr2018}. The vertical shear instability is explicitly not studied in our work, as we are exclusively considering locally-isothermal scenarios. 

Further, thermally driven (subcritical) baroclinic instabilities \citep{Klahr2003, Petersen2007a, Petersen2007b, Klahr2018} and its linear phase the convective overstability \citep{Klahr_2014, Lyra2014, Latter2016} are based on radially perturbed gas parcels carrying entropy and have previously been shown to affect the particle mid-plane layer. There, inward moving gas radiatively thermalizes with its environment, which leads to buoyancy oscillations that in turn transport entropy outward. This mechanism was shown to form vortices that can trap particles and thus significantly alter the particle distribution in the mid-plane layer \citep{Manger2018, Klahr2018}.

The gas vertical shear instability, the subcritical baroclinic instability, and the convective overstability, as well as others not yet mentioned, like the magneto-rotational instability \citep[see e.g.][]{Balbus1991, BalbusHawley1998, Davis2010, Bai2011, Bethune2016}, or zombie vortex instabilies \citep{Marcus2015, Marcus2016, Umurhan2016, Lesur2016} can introduce additional turbulence \citep{Klahr2018}, which as shown by \cite{Umurhan2019} reduces growth rates of SI modes and thus may render particle clumping and planetesimal formation via SI-induced over-densities more challenging.

Thus, while the diffusion-limited collapse criterion from \citet{klahr_schreiber_2015, Klahr2019_criterion} and Eq.~\eqref{eq:collapse_criterion_with_Q} is generally applicable for gravitationally collapse of over-dense clumps, the herein presented numerical verfication thereof and specifically the expression in Eq.~\eqref{eq:collapse_in_dead_zones} is most appropriate for protoplanetary disk dead zones.  In particular, our criteria apply when $\delta \gtrsim \alpha$, where the level of turbulence generated by other processes is given by the commonly-used parameter $\alpha$ \citep{Shakura} such that the turbulent velocity dispersion of the gas is $v_\mathrm{turb} = \alpha^{1/2}c_\mathrm{s}$ \citep[e.g.,][]{YoudinLithwick2007,Rosenthal2018}.  For the simulations in this paper, $\delta \sim 10^{-7}-10^{-5}$ depending on metallicity (see Figs.~\ref{fig:Ri_vs_z}, \ref{fig:vscale_vs_beta_and_Z}) and Stokes number after Eq.~\eqref{eq:particle_scale_height_from_deltaz}.

Recently, the DSHARP survey \citep[see e.g.,][]{Andrews2018} and likewise the Ophiuchus disk survey employing ALMA \citep[][]{Cieza2019} showed that rings are an abundant substructure in protoplanetary disks, which may hint at the existence of planets carving gaps in the gas disk, which leads to a pressure bump collecting the radially drifting dust particles \citep{Pinilla2012}. Recently, \citet{Stammler2019} were also able to explain the observed rings as a by-product of ongoing planetesimal formation. Regardless, it seems likely that pressure bumps constitute a common phenomena in protoplanetary disks \citep{Dullemond2018}, and one must therefore ask about the applicability of our collapse criterion therein. Pressure bumps lead to a local decrease in pressure gradient $\beta$. As a result, the relative dust-gas velocity decreases and KHI weakens leading to a thinner particle layer. As this effect is reflected in the proportionality of $Q_\mathrm{p}$ in $\beta$ in Eq.~\eqref{eq:collapse_in_dead_zones}, we expect pressure bumps to favor planetesimal formation, which agrees with findings by \citet{Dittrich2013} who investigated planetesimal formation in MRI-induced zonal flows and pressure bumps.

Lastly, we acknowledge that the evolution of formed clumps has not been studied in our work. This includes planetesimal growth, mergers and accretion \citep[e.g.,][]{Kokubo2012, SanSebastian2019, Liu2019, Johansen2019}, possible fragmentation \citep{Wakita2017, Gerbig2019}, migration \citep[e.g.,][]{Goldreich1979, MurrayClay2006, Kley2012}, pebble accretion \citep[e.g.,][]{Ormel2010, Bitsch2015, Rosenthal2019} and fragmenting and gravitoturbulent disks \citep{Gibbons2012, Booth2016, Baehr2019}.

\subsection{Outlook and conclusion}
\label{sect:outlook_conclusion}

Our results show that particle concentration by streaming instability and planetesimal formation through gravitational collapse are not equivalent. As such, our simulations were able to produce planetesimals for a metallicity of $Z = 0.01$ for $Q \lesssim 2$, even though no prior streaming instability clumping occurred, and likewise did not fragment for high $Q$-values regardless of significant prior SI-induced particle concentration.

Instead, the diffusion limited collapse criterion in Eq.~\eqref{eq:collapse_in_dead_zones} implies that whether or not planetesimals form depends only on disk mass via the Toomre parameter $Q$, the ratio of pressure gradient and metallicity $\beta/Z$ \citep[compare to][]{Sekiya2018}, and the local enhancement in dust-content $\epsilon$.

The herein presented Toomre-like formulation of the diffusion-limited collapse criterion also may be able to shed new light on the initial planetesimal mass function. Eq.~\eqref{eq:unstable_scales} predicts a range of length scales subject to non-linear gravitational instability, determined by the value of the right-hand side of Eq.~\eqref{eq:collapse_criterion_with_Q} (or Eq.~\eqref{eq:collapse_in_dead_zones}). By studying the dispersion relation in more detail, one can attach growth rates to each unstable scale. which can then be used to analytically inform an initial planetesimal mass function. 

In past numerical studies of the initial mass function, e.g. \citet{Simon2016, Schaefer2017, Nesvorny2019}, the enhancement $\epsilon$ is the only one of the quantities determining cloud collapse that is evolving in time. Therefore, the range of unstable scales and also initial planetesimal masses in past numerical studies may have been predominantly affected by parameter choices instead of characteristic properties of streaming and Kelvin-Helmholtz instability. This interpretation could explain why \citet{Simon2016} formed planetesimals when self-gravity was active from the start --- the collapse criterion in Eq.~\eqref{eq:collapse_criterion_with_Q} was already fulfilled by the initial condition --- and why they found that higher self-gravity parameters lead to more massive planetesimals: more scales are unstable for lower $Q$-values. Real disks however, are not expected to start off unstable to gravitational collapse, so one must ask how numerical simulations can inform initial planetesimal masses more physically. 

Our collapse criterion in Eq.~\eqref{eq:collapse_in_dead_zones} offers a possible solution to this conundrum.

In principle, there are four ways a system can evolve into fulfilling the collapse criterion. However, $Q$ and global pressure gradient $\beta$ in the absence of pressure bumps (see Sect.~\ref{sect:other_instabilities}) tend to be non-evolving. As such, softly turning on self-gravity as performed by \citet{Schaefer2017} does not necessarily reflect real conditions in protoplanetary disks, in particular since the strength of self-gravity does not explicitly affect SI growth rates. This is in contrast to the metallicity and Stokes number, which both highly influence streaming instability clumping \citep[e.g.,][]{Carrera2015}, but also inform the collapse criterion directly (for the metallicity) and implicitly via the enhancement parameter $\epsilon$.

Thus, to prevent the initial condition from collapsing, a simulation could be set up with a temporally increasing metallicity and having self-gravity turned on from the start. Then, planetesimals will only be formed once the collapse criterion is fulfilled. For high mass disks, we expect the mid-plane to start collapsing before the metallicity is high enough for streaming instability to start (compare to our $Z = 0.01$ simulation). This is of particular interest in the light of recent results by \citet{Powell2019} suggesting that disks may tend to be more massive than previously appreciated.

On the other hand, for low mass disks, we expect the onset of streaming instability to occur before the collapse criterion is fulfilled. In this case, streaming instability can locally increase particle densities, such that collapse occurs as soon as $\epsilon$ is high enough. A similar effect can be achieved by gradually increasing the Stokes number of super-particles. Both approaches mimic drift and growth processes in the sense that the abundance of particles with Stokes numbers appropriate for streaming instability increases over time.

Regardless of if the entire mid-plane or only locally enhanced regions collapse, we expect that, unless the growth timescale of the perturbation is shorter than its collapse timescale, marginally unstable modes will collapse first (i.e., for the right-hand side of Eq.~\eqref{eq:collapse_criterion_with_Q} equal to unity). This would prevent collapse of larger (or smaller) scales and dictate a single characteristic and dominant planetesimal size determined by the critical length scale in Eq.~\eqref{eq:critical_lengthscale} \citep[see][]{klahr_schreiber_2015, Klahr2019_criterion}, and therefore potentially lead to a initial planetesimal mass function that is qualitatively different to the power law seen in e.g., \citet{Johansen2015, Simon2016}.

To conclude, our analytical considerations and numerical simulations suggest that gravitational collapse of over-dense clouds does equally depend on metallicity, pressure gradient, Toomre $Q$ and potential local enhancements. These quantities should therefore be evaluated together, when assessing planetesimal formation. Streaming and Kelvin-Helmholtz instability are crucial in regulating local over-densities. Our collapse criterion remains to be tested for higher resolutions, different Stokes numbers and pressure gradients and for a larger range of metallicities.

\acknowledgments

KG thanks Orkan Umurhan, Paul Estrada, Til Birnstiel, Marco Vetter, Diana Powell, Mickey Rosenthal, John McCann, Eve Lee, Jake Simon and Andreas Schreiber for fruitful discussions. KG thanks UC Santa Cruz for hosting an extended visit. RMC acknowledges support from NSF CAREER grant number AST-1555385. Simulations were performed on the ISAAC cluster owned by the MPIA and hosted at the Max Planck Computing and Data Facility in Garching, Germany.

\software{The Pencil Code \citep{Brandenburg2002, Brandenburg2003}, Matplotlib \citep{Hunter2007}, Numpy \& Scipy \citep{Jones2001, Walt2011}.}


\vspace{2cm}

\appendix 

\section{On vertical and radial diffusivity driven by streaming and Kelvin-Helmholtz instability}

\label{sect:radial_vs_vertical_diff}

Diffusion plays a crucial role in limiting graviational collapse of a particle cloud on small scales. Throughout this paper, we used the vertical particle scale height as a proxy for vertical diffusivity to evaluate the collapse criterion in Eq.~\eqref{eq:collapse_criterion_with_Q} under the assumption of spherical symmetric diffusion. This procedure has the benefit of being easier to conduct compared to measurements of radial or azimuthal diffusivities, as neither collective drift nor Keplerian shear affect vertical particle velocities, and thus one can forgo the use of tracer particles \citep[which were utilized by e.g.,][]{Schreiber2018}.

However, as seen in 3D simulations of SI-driven turbulence by \citet{Johansen2007ApJ} as well as 2D simulations by \citet{Schreiber2018}, who compared the streaming instability in $x-z$ with its counterpart in $x-y$, radial and vertical diffusion driven by pure streaming instability tend to deviate by an order unity factor. Indeed, in our simulations, the radial component of the local root-mean-squared velocity of a local N-particle system measured via
\begin{align}
    v_\mathrm{rms,x} = \sqrt{\frac{1}{N}\sum_{i=1}^N|v_{x,i}- \langle v_x \rangle_N|^2}
\end{align}tends to be larger than the vertical component by up to one order of magnitude. This is in line with findings by \citet{Schreiber2018thesis}, even though they do not include vertical gravity and hence neither Kelvin-Helmholtz instability. Still, non-spherically-symmetric diffusivities should lead to an asymmetric gravitational collapse. In particular, an over-dense cloud may collapse faster or first in the vertical direction, thus potentially affecting planetesimal properties. We defer the detailed analysis of non-spherically-symmetric cloud collapse, that is not only including vertical but also radial and azimuthal diffusivity, to future work.

\section{Numerical robustness}

\subsection{Initial and boundary conditions}

\label{sect:initial_scale_height_numerical_test}

\begin{figure}[htp]
\includegraphics[width = \linewidth]{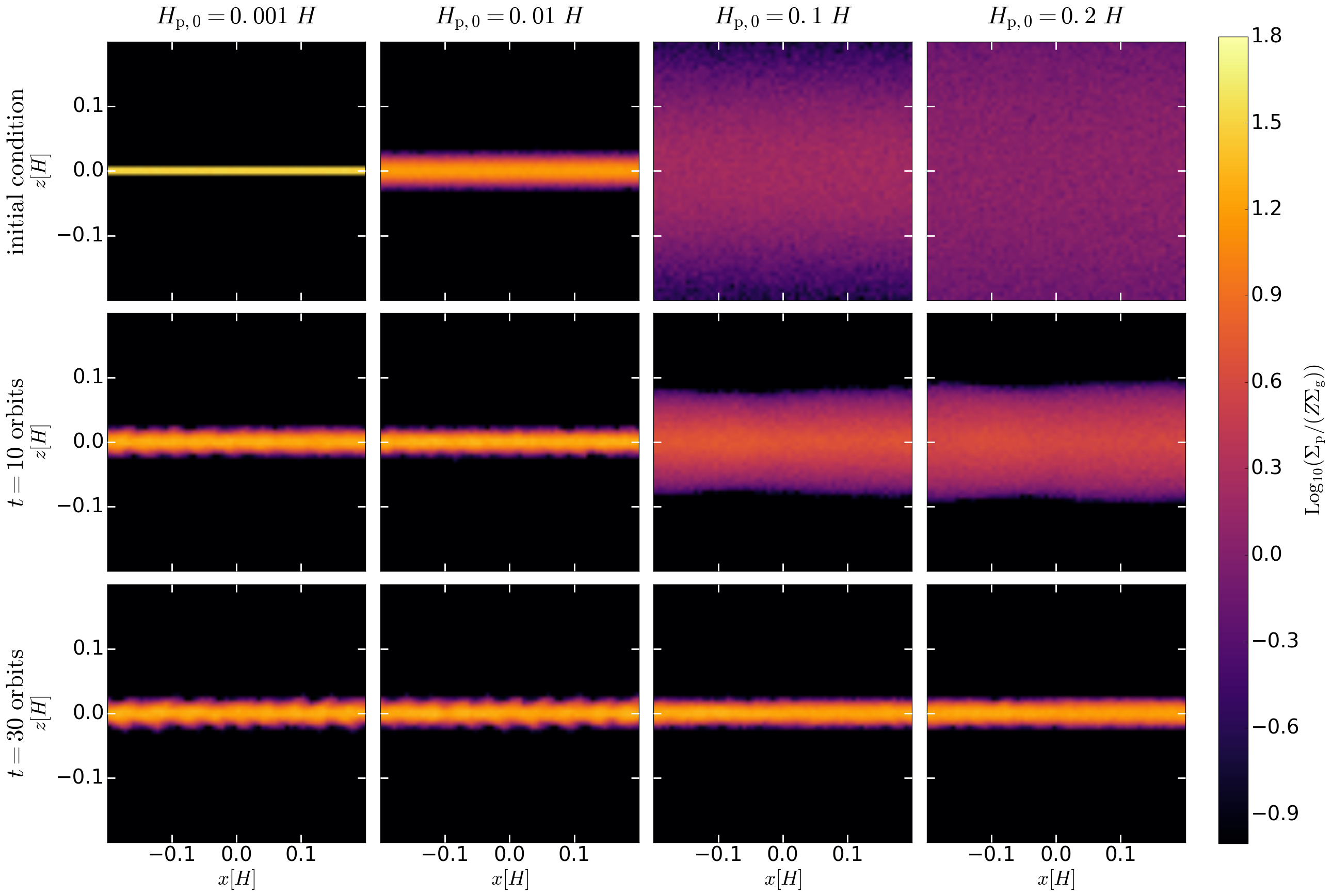}
\caption{Azimuthally integrated particle densities for the four settling simulations with $N_{x,y,z} = 64, \beta = 0.1, Z = 0.02, \mathrm{St} = 0.01$, but  different inital particle scale heights. After 30 orbits, all simulations have settled approximately to the expected scale $z_\mathrm{max}$ in Eq.~\eqref{eq:zmax}.\label{fig:initial_cond}}
\end{figure}

To verify that the vertical extent of the particle layer is independent of initial dust distribution, we performed particle settling tests with varying widths of the initial Gaussian distribution. For all simulations, we chose $N_{x,y,z} = 64, Z = 0.02$, $\mathrm{St} = 0.01$, and $\beta = 0.1$. We therefore expect a settling time of $\tau_\mathrm{set} \approx 15$ orbits (see Eq.~\eqref{eq:settling_time}). The result of this test is shown in Fig.~\ref{fig:initial_cond}. If particles are initialized below the expected extent, which for our choice of parameters is $z_\mathrm{max} \approx 0.25 H$ after Eq.~\eqref{eq:zmax}, they are excited to the expected scale, whereas they settle to $z_\mathrm{max}$ if they are initialized above this scale. We note, that the bottom left two panels in Fig.~\ref{fig:initial_cond} show more turbulent features than the bottom right panel, because here, the flow was initialized with a sub-critical Richardson number and the Kelvin-Helmholtz instability had to become active sooner to stir particles up.

We conclude that the initial particle scale height does not affect the final particle scale height, and $H_\mathrm{p,0}$ can in principle be chosen arbitrarily. In practice, for simulations with $\mathrm{St} = 0.005$, we initialize particles with a scale height close to the expected value in Eq.~\eqref{eq:zmax}, in order to minimize computational time. For $\mathrm{St} = 0.2$, the settling time in Eq.~\eqref{eq:settling_time} is very short and we can choose a higher initial height (also see Tab.~\ref{tab:simulations}).  

\subsection{Resolution and numerical convergence}

\label{sect:convergence}

In this paper, we present simulations with a resolution of 64x64x64 and 128x128x128 (see Tab.~\ref{tab:simulations}). In particular the lower resolution is not sufficient to resolve the fastest growing streaming instability modes which are always on the smallest scales. Still, due to being numerically inexpensive it allowed us to conduct a comparatively extensive parameter study. Moreover, the diffusion-limited collapse criterion presented in Sect.~\ref{sec:collapse_crit} and by \citet{Klahr2019_criterion} suggests that not the fastest growing SI wave-mode, but the critical radius in Eq.~\eqref{eq:critical_lengthscale} needs to be resolved to accurately present gravitational collapse and planetesimal formation. The fact that the low resolution fiducial run including self gravity required a lower $Q$-value for fragmentation to occur than the corresponding high resolution simulation (see Figs.~\ref{fig:Selfgrav_rhopmax} and \ref{fig:planetesimals}), suggests that the low resolution simulations were not in fact able to sufficiently resolve $r_\mathrm{crit}$.  Since, we do not test our criterion for a resolution exceeding $N_{x,y,z} = 128$, we can not be sure that our high resolution simulations are converged. Still, our estimate for the critical radius of $r_\mathrm{crit} \approx 0.006 H$ covers two grid cells for $N_{x,y,z} = 128$. It therefore stands those simulations are if not fully converged, close to numerical convergence.

We note that due to the strong spatial and temporal fluctuations in particle concentration inherit to the nonlinear saturation of streaming instability, numerical convergence is challenging to test by solely evaluating fragmentation. For example, the evolution of the maximum particle density shown in Fig.~\ref{fig:Selfgrav_rhopmax} before self-gravity is turned on is qualitatively similar for the two fiducial runs. Due these challenges and the large computational cost of performing a comprehensive convergence test, we leave this exercise for future work.

\bibliography{bibliography}{}

\begin{thebibliography}{}
\expandafter\ifx\csname natexlab\endcsname\relax\def\natexlab#1{#1}\fi
\providecommand{\url}[1]{\href{#1}{#1}}
\providecommand{\dodoi}[1]{doi:~\href{http://doi.org/#1}{\nolinkurl{#1}}}
\providecommand{\doeprint}[1]{\href{http://ascl.net/#1}{\nolinkurl{http://ascl.net/#1}}}
\providecommand{\doarXiv}[1]{\href{https://arxiv.org/abs/#1}{\nolinkurl{https://arxiv.org/abs/#1}}}

\bibitem[{{Abod} {et~al.}(2018){Abod}, {Simon}, {Li}, {Armitage}, {Youdin}, \&
  {Kretke}}]{Abod2018}
{Abod}, C.~P., {Simon}, J.~B., {Li}, R., {et~al.} 2018, arXiv e-prints,
  arXiv:1810.10018.
\newblock \doarXiv{1810.10018}

\bibitem[{Andrews {et~al.}(2009)Andrews, Wilner, Hughes, Qi, \&
  Dullemond}]{Andrews2009}
Andrews, S.~M., Wilner, D.~J., Hughes, A.~M., Qi, C., \& Dullemond, C.~P. 2009,
  The Astrophysical Journal, 700, 1502, \dodoi{10.1088/0004-637x/700/2/1502}

\bibitem[{{Andrews} {et~al.}(2018){Andrews}, {Huang}, {P{\'e}rez}, {Isella},
  {Dullemond}, {Kurtovic}, {Guzm{\'a}n}, {Carpenter}, {Wilner}, {Zhang}, {Zhu},
  {Birnstiel}, {Bai}, {Benisty}, {Hughes}, {{\"O}berg}, \&
  {Ricci}}]{Andrews2018}
{Andrews}, S.~M., {Huang}, J., {P{\'e}rez}, L.~M., {et~al.} 2018, \apjl, 869,
  L41, \dodoi{10.3847/2041-8213/aaf741}

\bibitem[{{Arlt} \& {Urpin}(2004)}]{Arlt2004}
{Arlt}, R., \& {Urpin}, V. 2004, A\&A, 426, 755,
  \dodoi{10.1051/0004-6361:20035896}

\bibitem[{Baehr \& Klahr(2019)}]{Baehr2019}
Baehr, H., \& Klahr, H. 2019, The Astrophysical Journal, 881, 162,
  \dodoi{10.3847/1538-4357/ab2f85}

\bibitem[{Baehr {et~al.}(2017)Baehr, Klahr, \& Kratter}]{Baehr2017}
Baehr, H., Klahr, H., \& Kratter, K.~M. 2017, The Astrophysical Journal, 848,
  40, \dodoi{10.3847/1538-4357/aa8a66}

\bibitem[{{Bai}(2011)}]{Bai2011}
{Bai}, X.-N. 2011, \apj, 739, 50, \dodoi{10.1088/0004-637X/739/1/50}

\bibitem[{{Bai} \&
  {Stone}(2010{\natexlab{a}})}]{Bai2010_ImplicationsPlanetesimals}
{Bai}, X.-N., \& {Stone}, J.~M. 2010{\natexlab{a}}, \apj, 722, 1437,
  \dodoi{10.1088/0004-637X/722/2/1437}

\bibitem[{{Bai} \& {Stone}(2010{\natexlab{b}})}]{Bai2010_pressuregradient}
---. 2010{\natexlab{b}}, \apj, 722, L220, \dodoi{10.1088/2041-8205/722/2/L220}

\bibitem[{{Balbus} \& {Hawley}(1991)}]{Balbus1991}
{Balbus}, S.~A., \& {Hawley}, J.~F. 1991, \apj, 376, 214,
  \dodoi{10.1086/170270}

\bibitem[{{Balbus} \& {Hawley}(1992)}]{Balbus1992}
---. 1992, \apj, 400, 610, \dodoi{10.1086/172022}

\bibitem[{Balbus \& Hawley(1998)}]{BalbusHawley1998}
Balbus, S.~A., \& Hawley, J.~F. 1998, Rev. Mod. Phys., 70, 1,
  \dodoi{10.1103/RevModPhys.70.1}

\bibitem[{Barker \& Latter(2015)}]{Barker2015}
Barker, A.~J., \& Latter, H.~N. 2015, Monthly Notices of the Royal Astronomical
  Society, 450, 21, \dodoi{10.1093/mnras/stv640}

\bibitem[{Belan {et~al.}(2014)Belan, Fouxon, \& Falkovich}]{Belan_2014}
Belan, S., Fouxon, I., \& Falkovich, G. 2014, Physical Review Letters, 112,
  \dodoi{10.1103/physrevlett.112.234502}

\bibitem[{{B{\'e}thune} {et~al.}(2016){B{\'e}thune}, {Lesur}, \&
  {Ferreira}}]{Bethune2016}
{B{\'e}thune}, W., {Lesur}, G., \& {Ferreira}, J. 2016, \aap, 589, A87,
  \dodoi{10.1051/0004-6361/201527874}

\bibitem[{{Birnstiel} {et~al.}(2010){Birnstiel}, {Dullemond}, \&
  {Brauer}}]{Birnstiel2010}
{Birnstiel}, T., {Dullemond}, C.~P., \& {Brauer}, F. 2010, \aap, 513, A79,
  \dodoi{10.1051/0004-6361/200913731}

\bibitem[{{Birnstiel} {et~al.}(2016){Birnstiel}, {Fang}, \&
  {Johansen}}]{Birnstiel2016}
{Birnstiel}, T., {Fang}, M., \& {Johansen}, A. 2016, \ssr, 205, 41,
  \dodoi{10.1007/s11214-016-0256-1}

\bibitem[{{Birnstiel} {et~al.}(2012){Birnstiel}, {Klahr}, \&
  {Ercolano}}]{Birnstiel2012}
{Birnstiel}, T., {Klahr}, H., \& {Ercolano}, B. 2012, \aap, 539, A148,
  \dodoi{10.1051/0004-6361/201118136}

\bibitem[{{Birnstiel} {et~al.}(2011){Birnstiel}, {Ormel}, \&
  {Dullemond}}]{Birnstiel2011}
{Birnstiel}, T., {Ormel}, C.~W., \& {Dullemond}, C.~P. 2011, \aap, 525, A11,
  \dodoi{10.1051/0004-6361/201015228}

\bibitem[{{Bitsch} {et~al.}(2015){Bitsch}, {Lambrechts}, \&
  {Johansen}}]{Bitsch2015}
{Bitsch}, B., {Lambrechts}, M., \& {Johansen}, A. 2015, \aap, 582, A112,
  \dodoi{10.1051/0004-6361/201526463}

\bibitem[{Booth \& Clarke(2016)}]{Booth2016}
Booth, R.~A., \& Clarke, C.~J. 2016, Monthly Notices of the Royal Astronomical
  Society, 458, 2676, \dodoi{10.1093/mnras/stw488}

\bibitem[{Brandenburg(2003)}]{Brandenburg2003}
Brandenburg, A. 2003, The Fluid Mechanics of Astrophysics and Geophysics,
  269–344, \dodoi{10.1201/9780203493137.ch9}

\bibitem[{{Brandenburg} \& {Dobler}(2002)}]{Brandenburg2002}
{Brandenburg}, A., \& {Dobler}, W. 2002, Computer Physics Communications, 147,
  471, \dodoi{10.1016/S0010-4655(02)00334-X}

\bibitem[{{Brandenburg} {et~al.}(1995){Brandenburg}, {Nordlund}, {Stein}, \&
  {Torkelsson}}]{Brandenburg1995}
{Brandenburg}, A., {Nordlund}, A., {Stein}, R.~F., \& {Torkelsson}, U. 1995,
  \apj, 446, 741, \dodoi{10.1086/175831}

\bibitem[{{Brauer} {et~al.}(2008){Brauer}, {Dullemond}, \&
  {Henning}}]{Brauer2008}
{Brauer}, F., {Dullemond}, C.~P., \& {Henning}, T. 2008, \aap, 480, 859,
  \dodoi{10.1051/0004-6361:20077759}

\bibitem[{{Caporaloni} {et~al.}(1975){Caporaloni}, {Tampieri}, {Trombetti}, \&
  {Vittori}}]{Caporaloni1975}
{Caporaloni}, M., {Tampieri}, F., {Trombetti}, F., \& {Vittori}, O. 1975,
  Journal of Atmospheric Sciences, 32, 565,
  \dodoi{10.1175/1520-0469(1975)032<0565:TOPINA>2.0.CO;2}

\bibitem[{{Carrera} {et~al.}(2015){Carrera}, {Johansen}, \&
  {Davies}}]{Carrera2015}
{Carrera}, D., {Johansen}, A., \& {Davies}, M.~B. 2015, \aap, 579, A43,
  \dodoi{10.1051/0004-6361/201425120}

\bibitem[{{Chandrasekhar}(1961)}]{Chandrasekhar1961}
{Chandrasekhar}, S. 1961, {Hydrodynamic and hydromagnetic stability} (Dover
  Publications)

\bibitem[{{Chiang}(2008)}]{Chiang2008}
{Chiang}, E. 2008, \apj, 675, 1549, \dodoi{10.1086/527354}

\bibitem[{{Chiang} {et~al.}(2014){Chiang}, {Murray-Clay}, \&
  {Shi}}]{Chiang2014}
{Chiang}, E., {Murray-Clay}, R., \& {Shi}, J.-M. 2014, in IAU Symposium, Vol.
  299, Exploring the Formation and Evolution of Planetary Systems, ed.
  M.~{Booth}, B.~C. {Matthews}, \& J.~R. {Graham}, 136--139,
  \dodoi{10.1017/S1743921313008119}

\bibitem[{{Chiang} \& {Youdin}(2010)}]{Chiang2010}
{Chiang}, E., \& {Youdin}, A.~N. 2010, Annual Review of Earth and Planetary
  Sciences, 38, 493, \dodoi{10.1146/annurev-earth-040809-152513}

\bibitem[{{Cieza} {et~al.}(2019){Cieza}, {Ru{\'\i}z-Rodr{\'\i}guez}, {Hales},
  {Casassus}, {P{\'e}rez}, {Gonzalez-Ruilova}, {C{\'a}novas}, {Williams},
  {Zurlo}, {Ansdell}, {Avenhaus}, {Bayo}, {Bertrang}, {Christiaens}, {Dent},
  {Ferrero}, {Gamen}, {Olofsson}, {Orcajo}, {Pe{\~n}a Ram{\'\i}rez},
  {Principe}, {Schreiber}, \& {van der Plas}}]{Cieza2019}
{Cieza}, L.~A., {Ru{\'\i}z-Rodr{\'\i}guez}, D., {Hales}, A., {et~al.} 2019,
  \mnras, 482, 698, \dodoi{10.1093/mnras/sty2653}

\bibitem[{{Davis} {et~al.}(2010){Davis}, {Stone}, \& {Pessah}}]{Davis2010}
{Davis}, S.~W., {Stone}, J.~M., \& {Pessah}, M.~E. 2010, \apj, 713, 52,
  \dodoi{10.1088/0004-637X/713/1/52}

\bibitem[{{Dittrich} {et~al.}(2013){Dittrich}, {Klahr}, \&
  {Johansen}}]{Dittrich2013}
{Dittrich}, K., {Klahr}, H., \& {Johansen}, A. 2013, \apj, 763, 117,
  \dodoi{10.1088/0004-637X/763/2/117}

\bibitem[{{Dr{\k{a}}{\.z}kowska} {et~al.}(2016){Dr{\k{a}}{\.z}kowska},
  {Alibert}, \& {Moore}}]{Drazkowska2016}
{Dr{\k{a}}{\.z}kowska}, J., {Alibert}, Y., \& {Moore}, B. 2016, \aap, 594,
  A105, \dodoi{10.1051/0004-6361/201628983}

\bibitem[{{Dullemond} {et~al.}(2018){Dullemond}, {Birnstiel}, {Huang},
  {Kurtovic}, {Andrews}, {Guzm{\'a}n}, {P{\'e}rez}, {Isella}, {Zhu}, {Benisty},
  {Wilner}, {Bai}, {Carpenter}, {Zhang}, \& {Ricci}}]{Dullemond2018}
{Dullemond}, C.~P., {Birnstiel}, T., {Huang}, J., {et~al.} 2018, \apjl, 869,
  L46, \dodoi{10.3847/2041-8213/aaf742}

\bibitem[{{Epstein}(1924)}]{Epstein1924}
{Epstein}, P.~S. 1924, Physical Review, 23, 710, \dodoi{10.1103/PhysRev.23.710}

\bibitem[{{Gammie}(2001)}]{Gammie2001}
{Gammie}, C.~F. 2001, \apj, 553, 174, \dodoi{10.1086/320631}

\bibitem[{Gerbig {et~al.}(2019)Gerbig, Lenz, \& Klahr}]{Gerbig2019}
Gerbig, K., Lenz, C.~T., \& Klahr, H. 2019, \aap, 629, A116,
  \dodoi{10.1051/0004-6361/201935278}

\bibitem[{Gibbons {et~al.}(2012)Gibbons, Rice, \& Mamatsashvili}]{Gibbons2012}
Gibbons, P.~G., Rice, W. K.~M., \& Mamatsashvili, G.~R. 2012, Monthly Notices
  of the Royal Astronomical Society, 426, 1444,
  \dodoi{10.1111/j.1365-2966.2012.21731.x}

\bibitem[{{Goldreich} \& {Lynden-Bell}(1965)}]{Goldreich1965}
{Goldreich}, P., \& {Lynden-Bell}, D. 1965, \mnras, 130, 125,
  \dodoi{10.1093/mnras/130.2.125}

\bibitem[{{Goldreich} \& {Tremaine}(1979)}]{Goldreich1979}
{Goldreich}, P., \& {Tremaine}, S. 1979, \apj, 233, 857, \dodoi{10.1086/157448}

\bibitem[{{Goldreich} \& {Ward}(1973)}]{Goldreich1973}
{Goldreich}, P., \& {Ward}, W.~R. 1973, \apj, 183, 1051, \dodoi{10.1086/152291}

\bibitem[{{G{\'o}mez} \& {Ostriker}(2005)}]{Gomez2005}
{G{\'o}mez}, G.~C., \& {Ostriker}, E.~C. 2005, \apj, 630, 1093,
  \dodoi{10.1086/432086}

\bibitem[{{Gorti} {et~al.}(2015){Gorti}, {Hollenbach}, \&
  {Dullemond}}]{Gorti2015}
{Gorti}, U., {Hollenbach}, D., \& {Dullemond}, C.~P. 2015, \apj, 804, 29,
  \dodoi{10.1088/0004-637X/804/1/29}

\bibitem[{{Hayashi}(1981)}]{Hayashi1981}
{Hayashi}, C. 1981, Progress of Theoretical Physics Supplement, 70, 35,
  \dodoi{10.1143/PTPS.70.35}

\bibitem[{Hockney \& Eastwood(1988)}]{hockney1988computer}
Hockney, R., \& Eastwood, J. 1988, Computer Simulation Using Particles (CRC
  Press).
\newblock \url{https://books.google.com/books?id=nTOFkmnCQuIC}

\bibitem[{Howard \& Maslowe(1973)}]{Howard1973}
Howard, L.~N., \& Maslowe, S.~A. 1973, Boundary-Layer Meteorology, 4, 511,
  \dodoi{10.1007/BF02265252}

\bibitem[{{Hunter}(2007)}]{Hunter2007}
{Hunter}, J.~D. 2007, Computing in Science Engineering, 9, 90,
  \dodoi{10.1109/MCSE.2007.55}

\bibitem[{{Jacquet} {et~al.}(2011){Jacquet}, {Balbus}, \&
  {Latter}}]{Jacquet2011}
{Jacquet}, E., {Balbus}, S., \& {Latter}, H. 2011, \mnras, 415, 3591,
  \dodoi{10.1111/j.1365-2966.2011.18971.x}

\bibitem[{{Johansen} \& {Bitsch}(2019)}]{Johansen2019}
{Johansen}, A., \& {Bitsch}, B. 2019, arXiv e-prints, arXiv:1909.10429.
\newblock \doarXiv{1909.10429}

\bibitem[{{Johansen} {et~al.}(2006){Johansen}, {Henning}, \&
  {Klahr}}]{Johansen2006KHI}
{Johansen}, A., {Henning}, T., \& {Klahr}, H. 2006, \apj, 643, 1219,
  \dodoi{10.1086/502968}

\bibitem[{Johansen {et~al.}(2006{\natexlab{a}})Johansen, Klahr, \&
  Henning}]{Johansen2006}
Johansen, A., Klahr, H., \& Henning, T. 2006{\natexlab{a}}, The Astrophysical
  Journal, 636, 1121, \dodoi{10.1086/498078}

\bibitem[{Johansen {et~al.}(2006{\natexlab{b}})Johansen, Klahr, \&
  Mee}]{Johansen2006_diffusion}
Johansen, A., Klahr, H., \& Mee, A.~J. 2006{\natexlab{b}}, Mon. Not. Roy.
  Astron. Soc., 370, L71, \dodoi{10.1111/j.1745-3933.2006.00191.x}

\bibitem[{{Johansen} {et~al.}(2015){Johansen}, {Mac Low}, {Lacerda}, \&
  {Bizzarro}}]{Johansen2015}
{Johansen}, A., {Mac Low}, M.-M., {Lacerda}, P., \& {Bizzarro}, M. 2015,
  Science Advances, 1, 1500109, \dodoi{10.1126/sciadv.1500109}

\bibitem[{{Johansen} {et~al.}(2007){Johansen}, {Oishi}, {Mac Low}, {Klahr},
  {Henning}, \& {Youdin}}]{Johansen_2007nature}
{Johansen}, A., {Oishi}, J.~S., {Mac Low}, M.-M., {et~al.} 2007, \nat, 448,
  1022, \dodoi{10.1038/nature06086}

\bibitem[{Johansen \& Youdin(2007)}]{Johansen2007ApJ}
Johansen, A., \& Youdin, A. 2007, The Astrophysical Journal, 662, 627,
  \dodoi{10.1086/516730}

\bibitem[{{Johansen} {et~al.}(2009){Johansen}, {Youdin}, \& {Mac
  Low}}]{Johansen_2009}
{Johansen}, A., {Youdin}, A., \& {Mac Low}, M.-M. 2009, \apj, 704, L75,
  \dodoi{10.1088/0004-637X/704/2/L75}

\bibitem[{{Johansen} {et~al.}(2012){Johansen}, {Youdin}, \&
  {Lithwick}}]{Johansen2012}
{Johansen}, A., {Youdin}, A.~N., \& {Lithwick}, Y. 2012, \aap, 537, A125,
  \dodoi{10.1051/0004-6361/201117701}

\bibitem[{Jones {et~al.}(2001--)Jones, Oliphant, Peterson,
  {et~al.}}]{Jones2001}
Jones, E., Oliphant, T., Peterson, P., {et~al.} 2001--, {SciPy}: Open source
  scientific tools for {Python}.
\newblock \url{http://www.scipy.org/}

\bibitem[{{Kataoka} {et~al.}(2013){Kataoka}, {Tanaka}, {Okuzumi}, \&
  {Wada}}]{Kataoka2013}
{Kataoka}, A., {Tanaka}, H., {Okuzumi}, S., \& {Wada}, K. 2013, \aap, 557, L4,
  \dodoi{10.1051/0004-6361/201322151}

\bibitem[{Klahr \& Hubbard(2014)}]{Klahr_2014}
Klahr, H., \& Hubbard, A. 2014, The Astrophysical Journal, 788, 21,
  \dodoi{10.1088/0004-637x/788/1/21}

\bibitem[{Klahr {et~al.}(2018)Klahr, Pfeil, \& Schreiber}]{Klahr2018}
Klahr, H., Pfeil, T., \& Schreiber, A. 2018, Instabilities and Flow Structures
  in Protoplanetary Disks: Setting the Stage for Planetesimal Formation (Cham:
  Springer International Publishing), 2251--2286,
  \dodoi{10.1007/978-3-319-55333-7_138}

\bibitem[{Klahr \& Schreiber(2015)}]{klahr_schreiber_2015}
Klahr, H., \& Schreiber, A. 2015, Proceedings of the International Astronomical
  Union, 10, 1–8, \dodoi{10.1017/S1743921315010406}

\bibitem[{{Klahr} \& {Schreiber}(2019)}]{Klahr2019_criterion}
{Klahr}, H., \& {Schreiber}, A. 2019, Submitted to ApJ

\bibitem[{Klahr \& Bodenheimer(2003)}]{Klahr2003}
Klahr, H.~H., \& Bodenheimer, P. 2003, The Astrophysical Journal, 582, 869,
  \dodoi{10.1086/344743}

\bibitem[{{Kley} \& {Nelson}(2012)}]{Kley2012}
{Kley}, W., \& {Nelson}, R.~P. 2012, \araa, 50, 211,
  \dodoi{10.1146/annurev-astro-081811-125523}

\bibitem[{{Kokubo} \& {Ida}(2012)}]{Kokubo2012}
{Kokubo}, E., \& {Ida}, S. 2012, Progress of Theoretical and Experimental
  Physics, 2012, 01A308, \dodoi{10.1093/ptep/pts032}

\bibitem[{{Krapp} {et~al.}(2019){Krapp}, {Ben{\'\i}tez-Llambay}, {Gressel}, \&
  {Pessah}}]{Krapp2019}
{Krapp}, L., {Ben{\'\i}tez-Llambay}, P., {Gressel}, O., \& {Pessah}, M.~E.
  2019, \apjl, 878, L30, \dodoi{10.3847/2041-8213/ab2596}

\bibitem[{{Kratter} \& {Murray-Clay}(2011)}]{Kratter2011}
{Kratter}, K.~M., \& {Murray-Clay}, R.~A. 2011, \apj, 740, 1,
  \dodoi{10.1088/0004-637X/740/1/1}

\bibitem[{{Latter}(2016)}]{Latter2016}
{Latter}, H.~N. 2016, \mnras, 455, 2608, \dodoi{10.1093/mnras/stv2449}

\bibitem[{{Lenz} {et~al.}(2019){Lenz}, {Klahr}, \& {Birnstiel}}]{Lenz2019}
{Lenz}, C.~T., {Klahr}, H., \& {Birnstiel}, T. 2019, \apj, 874, 36,
  \dodoi{10.3847/1538-4357/ab05d9}

\bibitem[{{Lesur} \& {Latter}(2016)}]{Lesur2016}
{Lesur}, G. R.~J., \& {Latter}, H. 2016, \mnras, 462, 4549,
  \dodoi{10.1093/mnras/stw2172}

\bibitem[{{Li} {et~al.}(2003){Li}, {Goodman}, \& {Narayan}}]{Li2003}
{Li}, L.-X., {Goodman}, J., \& {Narayan}, R. 2003, \apj, 593, 980,
  \dodoi{10.1086/376695}

\bibitem[{Li {et~al.}(2018)Li, Youdin, \& Simon}]{Li2018}
Li, R., Youdin, A.~N., \& Simon, J.~B. 2018, The Astrophysical Journal, 862,
  14, \dodoi{10.3847/1538-4357/aaca99}

\bibitem[{{Li} {et~al.}(2019){Li}, {Youdin}, \& {Simon}}]{Li2019}
{Li}, R., {Youdin}, A.~N., \& {Simon}, J.~B. 2019, \apj, 885, 69,
  \dodoi{10.3847/1538-4357/ab480d}

\bibitem[{{Lin} \& {Youdin}(2015)}]{Lin2015}
{Lin}, M.-K., \& {Youdin}, A.~N. 2015, \apj, 811, 17,
  \dodoi{10.1088/0004-637X/811/1/17}

\bibitem[{{Liu} {et~al.}(2019){Liu}, {Ormel}, \& {Johansen}}]{Liu2019}
{Liu}, B., {Ormel}, C.~W., \& {Johansen}, A. 2019, \aap, 624, A114,
  \dodoi{10.1051/0004-6361/201834174}

\bibitem[{{Lyra}(2014)}]{Lyra2014}
{Lyra}, W. 2014, \apj, 789, 77, \dodoi{10.1088/0004-637X/789/1/77}

\bibitem[{{Manger} \& {Klahr}(2018)}]{Manger2018}
{Manger}, N., \& {Klahr}, H. 2018, \mnras, 480, 2125,
  \dodoi{10.1093/mnras/sty1909}

\bibitem[{{Marcus} {et~al.}(2016){Marcus}, {Pei}, {Jiang}, \&
  {Barranco}}]{Marcus2016}
{Marcus}, P.~S., {Pei}, S., {Jiang}, C.-H., \& {Barranco}, J.~A. 2016, \apj,
  833, 148, \dodoi{10.3847/1538-4357/833/2/148}

\bibitem[{{Marcus} {et~al.}(2015){Marcus}, {Pei}, {Jiang}, {Barranco},
  {Hassanzadeh}, \& {Lecoanet}}]{Marcus2015}
{Marcus}, P.~S., {Pei}, S., {Jiang}, C.-H., {et~al.} 2015, \apj, 808, 87,
  \dodoi{10.1088/0004-637X/808/1/87}

\bibitem[{{Murray-Clay} \& {Chiang}(2006)}]{MurrayClay2006}
{Murray-Clay}, R.~A., \& {Chiang}, E.~I. 2006, \apj, 651, 1194,
  \dodoi{10.1086/507514}

\bibitem[{{Nakagawa} {et~al.}(1986){Nakagawa}, {Sekiya}, \&
  {Hayashi}}]{Nakagawa1986}
{Nakagawa}, Y., {Sekiya}, M., \& {Hayashi}, C. 1986, \icarus, 67, 375,
  \dodoi{10.1016/0019-1035(86)90121-1}

\bibitem[{{Nelson} {et~al.}(2013){Nelson}, {Gressel}, \&
  {Umurhan}}]{Nelson2013}
{Nelson}, R.~P., {Gressel}, O., \& {Umurhan}, O.~M. 2013, \mnras, 435, 2610,
  \dodoi{10.1093/mnras/stt1475}

\bibitem[{{Nesvorny} {et~al.}(2019){Nesvorny}, {Li}, {Youdin}, {Simon}, \&
  {Grundy}}]{Nesvorny2019}
{Nesvorny}, D., {Li}, R., {Youdin}, A.~N., {Simon}, J.~B., \& {Grundy}, W.~M.
  2019, arXiv e-prints, arXiv:1906.11344.
\newblock \doarXiv{1906.11344}

\bibitem[{{Okuzumi} {et~al.}(2012){Okuzumi}, {Tanaka}, {Kobayashi}, \&
  {Wada}}]{Okuzumi2012}
{Okuzumi}, S., {Tanaka}, H., {Kobayashi}, H., \& {Wada}, K. 2012, \apj, 752,
  106, \dodoi{10.1088/0004-637X/752/2/106}

\bibitem[{{Ormel} \& {Klahr}(2010)}]{Ormel2010}
{Ormel}, C.~W., \& {Klahr}, H.~H. 2010, A\&A, 520, A43,
  \dodoi{10.1051/0004-6361/201014903}

\bibitem[{{Petersen} {et~al.}(2007{\natexlab{a}}){Petersen}, {Julien}, \&
  {Stewart}}]{Petersen2007a}
{Petersen}, M.~R., {Julien}, K., \& {Stewart}, G.~R. 2007{\natexlab{a}}, \apj,
  658, 1236, \dodoi{10.1086/511513}

\bibitem[{{Petersen} {et~al.}(2007{\natexlab{b}}){Petersen}, {Stewart}, \&
  {Julien}}]{Petersen2007b}
{Petersen}, M.~R., {Stewart}, G.~R., \& {Julien}, K. 2007{\natexlab{b}}, \apj,
  658, 1252, \dodoi{10.1086/511523}

\bibitem[{{Pfeil} \& {Klahr}(2019)}]{Pfeil2019}
{Pfeil}, T., \& {Klahr}, H. 2019, \apj, 871, 150,
  \dodoi{10.3847/1538-4357/aaf962}

\bibitem[{{Pinilla} {et~al.}(2012){Pinilla}, {Benisty}, \&
  {Birnstiel}}]{Pinilla2012}
{Pinilla}, P., {Benisty}, M., \& {Birnstiel}, T. 2012, \aap, 545, A81,
  \dodoi{10.1051/0004-6361/201219315}

\bibitem[{{Powell} {et~al.}(2019){Powell}, {Murray-Clay}, {P{\'e}rez},
  {Schlichting}, \& {Rosenthal}}]{Powell2019}
{Powell}, D., {Murray-Clay}, R., {P{\'e}rez}, L.~M., {Schlichting}, H.~E., \&
  {Rosenthal}, M. 2019, \apj, 878, 116, \dodoi{10.3847/1538-4357/ab20ce}

\bibitem[{{Raettig} {et~al.}(2015){Raettig}, {Klahr}, \& {Lyra}}]{Raettig2015}
{Raettig}, N., {Klahr}, H., \& {Lyra}, W. 2015, \apj, 804, 35,
  \dodoi{10.1088/0004-637X/804/1/35}

\bibitem[{Reeks(2014)}]{Reeks2014}
Reeks, M.~W. 2014, Flow, Turbulence and Combustion, 92, 3,
  \dodoi{10.1007/s10494-013-9515-3}

\bibitem[{{Rice} {et~al.}(2006){Rice}, {Armitage}, {Wood}, \&
  {Lodato}}]{Rice2006}
{Rice}, W.~K.~M., {Armitage}, P.~J., {Wood}, K., \& {Lodato}, G. 2006, \mnras,
  373, 1619, \dodoi{10.1111/j.1365-2966.2006.11113.x}

\bibitem[{{Rosenthal} \& {Murray-Clay}(2019)}]{Rosenthal2019}
{Rosenthal}, M.~M., \& {Murray-Clay}, R.~A. 2019, arXiv e-prints,
  arXiv:1908.06991.
\newblock \doarXiv{1908.06991}

\bibitem[{{Rosenthal} {et~al.}(2018){Rosenthal}, {Murray-Clay}, {Perets}, \&
  {Wolansky}}]{Rosenthal2018}
{Rosenthal}, M.~M., {Murray-Clay}, R.~A., {Perets}, H.~B., \& {Wolansky}, N.
  2018, \apj, 861, 74, \dodoi{10.3847/1538-4357/aac4a1}

\bibitem[{Safronov(1969)}]{Safronov1969}
Safronov, V.~S. 1969, {Evolution of the protoplanetary cloud and formation of
  the earth and planets} (Nauka Press)

\bibitem[{{San Sebasti{\'a}n} {et~al.}(2019){San Sebasti{\'a}n}, {Guilera}, \&
  {Parisi}}]{SanSebastian2019}
{San Sebasti{\'a}n}, I.~L., {Guilera}, O.~M., \& {Parisi}, M.~G. 2019, \aap,
  625, A138, \dodoi{10.1051/0004-6361/201834168}

\bibitem[{{Sch{\"a}fer} {et~al.}(2017){Sch{\"a}fer}, {Yang}, \&
  {Johansen}}]{Schaefer2017}
{Sch{\"a}fer}, U., {Yang}, C.-C., \& {Johansen}, A. 2017, \aap, 597, A69,
  \dodoi{10.1051/0004-6361/201629561}

\bibitem[{{Schaffer} {et~al.}(2018){Schaffer}, {Yang}, \&
  {Johansen}}]{Schaffer2018}
{Schaffer}, N., {Yang}, C.-C., \& {Johansen}, A. 2018, \aap, 618, A75,
  \dodoi{10.1051/0004-6361/201832783}

\bibitem[{Schreiber(2018)}]{Schreiber2018thesis}
Schreiber, A. 2018, PhD thesis, Ruperto-Carola University Heidelberg

\bibitem[{Schreiber \& Klahr(2018)}]{Schreiber2018}
Schreiber, A., \& Klahr, H. 2018, The Astrophysical Journal, 861, 47.
\newblock \url{http://stacks.iop.org/0004-637X/861/i=1/a=47}

\bibitem[{Sekiya(1998)}]{Sekiya1998}
Sekiya, M. 1998, Icarus, 133, 298 ,
  \dodoi{https://doi.org/10.1006/icar.1998.5933}

\bibitem[{Sekiya \& Ishitsu(2000)}]{Sekiya2000}
Sekiya, M., \& Ishitsu, N. 2000, Earth, Planets and Space, 52, 517,
  \dodoi{10.1186/BF03351656}

\bibitem[{Sekiya \& Ishitsu(2001)}]{Sekiya2001}
---. 2001, Earth, Planets and Space, 53, 761, \dodoi{10.1186/BF03352404}

\bibitem[{Sekiya \& Onishi(2018)}]{Sekiya2018}
Sekiya, M., \& Onishi, I.~K. 2018, The Astrophysical Journal, 860, 140,
  \dodoi{10.3847/1538-4357/aac4a7}

\bibitem[{{Shakura} \& {Sunyaev}(1973)}]{Shakura}
{Shakura}, N.~I., \& {Sunyaev}, R.~A. 1973, \aap, 24, 337

\bibitem[{{Shi} \& {Chiang}(2013)}]{Shi2013}
{Shi}, J.-M., \& {Chiang}, E. 2013, \apj, 764, 20,
  \dodoi{10.1088/0004-637X/764/1/20}

\bibitem[{{Simon} {et~al.}(2016){Simon}, {Armitage}, {Li}, \&
  {Youdin}}]{Simon2016}
{Simon}, J.~B., {Armitage}, P.~J., {Li}, R., \& {Youdin}, A.~N. 2016, \apj,
  822, 55, \dodoi{10.3847/0004-637X/822/1/55}

\bibitem[{Simon {et~al.}(2017)Simon, Armitage, Youdin, \& Li}]{Simon2017}
Simon, J.~B., Armitage, P.~J., Youdin, A.~N., \& Li, R. 2017, The Astrophysical
  Journal, 847, L12, \dodoi{10.3847/2041-8213/aa8c79}

\bibitem[{{Squire} \& {Hopkins}(2018{\natexlab{a}})}]{Squire2018_RDIandSI}
{Squire}, J., \& {Hopkins}, P.~F. 2018{\natexlab{a}}, \mnras, 477, 5011,
  \dodoi{10.1093/mnras/sty854}

\bibitem[{{Squire} \& {Hopkins}(2018{\natexlab{b}})}]{Squire2018RDI}
---. 2018{\natexlab{b}}, \apj, 856, L15, \dodoi{10.3847/2041-8213/aab54d}

\bibitem[{{Stammler} {et~al.}(2019){Stammler}, {Dr{\k{a}}{\.z}kowska},
  {Birnstiel}, {Klahr}, {Dullemond}, \& {Andrews}}]{Stammler2019}
{Stammler}, S.~M., {Dr{\k{a}}{\.z}kowska}, J., {Birnstiel}, T., {et~al.} 2019,
  \apjl, 884, L5, \dodoi{10.3847/2041-8213/ab4423}

\bibitem[{{Taki} {et~al.}(2016){Taki}, {Fujimoto}, \& {Ida}}]{Taki2016}
{Taki}, T., {Fujimoto}, M., \& {Ida}, S. 2016, \aap, 591, A86,
  \dodoi{10.1051/0004-6361/201527732}

\bibitem[{{Toomre}(1964)}]{Toomre1964}
{Toomre}, A. 1964, \apj, 139, 1217, \dodoi{10.1086/147861}

\bibitem[{{Umurhan} {et~al.}(2019){Umurhan}, {Estrada}, \&
  {Cuzzi}}]{Umurhan2019}
{Umurhan}, O.~M., {Estrada}, P.~R., \& {Cuzzi}, J.~N. 2019, arXiv e-prints,
  arXiv:1906.05371.
\newblock \doarXiv{1906.05371}

\bibitem[{{Umurhan} \& {Regev}(2004)}]{Umurhan2004}
{Umurhan}, O.~M., \& {Regev}, O. 2004, \aap, 427, 855,
  \dodoi{10.1051/0004-6361:20040573}

\bibitem[{{Umurhan} {et~al.}(2016){Umurhan}, {Shariff}, \&
  {Cuzzi}}]{Umurhan2016}
{Umurhan}, O.~M., {Shariff}, K., \& {Cuzzi}, J.~N. 2016, \apj, 830, 95,
  \dodoi{10.3847/0004-637X/830/2/95}

\bibitem[{{Urpin} \& {Brandenburg}(1998)}]{Urpin1998}
{Urpin}, V., \& {Brandenburg}, A. 1998, \mnras, 294, 399,
  \dodoi{10.1046/j.1365-8711.1998.01118.x}

\bibitem[{{Wakita} {et~al.}(2017){Wakita}, {Matsumoto}, {Oshino}, \&
  {Hasegawa}}]{Wakita2017}
{Wakita}, S., {Matsumoto}, Y., {Oshino}, S., \& {Hasegawa}, Y. 2017, \apj, 834,
  125, \dodoi{10.3847/1538-4357/834/2/125}

\bibitem[{Walt {et~al.}(2011)Walt, Colbert, \& Varoquaux}]{Walt2011}
Walt, S. v.~d., Colbert, S.~C., \& Varoquaux, G. 2011, Computing in Science \&
  Engineering, 13, 22, \dodoi{10.1109/MCSE.2011.37}

\bibitem[{{Weidenschilling}(1977)}]{Weidenschilling1977MMSN}
{Weidenschilling}, S.~J. 1977, \apss, 51, 153, \dodoi{10.1007/BF00642464}

\bibitem[{{Weidenschilling}(1980)}]{Weidenschilling1980}
---. 1980, \icarus, 44, 172, \dodoi{10.1016/0019-1035(80)90064-0}

\bibitem[{{Weidenschilling} \& {Cuzzi}(1993)}]{Weidenschilling1993}
{Weidenschilling}, S.~J., \& {Cuzzi}, J.~N. 1993, in Protostars and Planets
  III, ed. E.~H. {Levy} \& J.~I. {Lunine}, 1031--1060

\bibitem[{{Yang} \& {Johansen}(2014)}]{Yang2014}
{Yang}, C.-C., \& {Johansen}, A. 2014, \apj, 792, 86,
  \dodoi{10.1088/0004-637X/792/2/86}

\bibitem[{{Yang} {et~al.}(2017){Yang}, {Johansen}, \& {Carrera}}]{Yang2017}
{Yang}, C.~C., {Johansen}, A., \& {Carrera}, D. 2017, \aap, 606, A80,
  \dodoi{10.1051/0004-6361/201630106}

\bibitem[{Yang \& Krumholz(2012)}]{Yang2012}
Yang, C.-C., \& Krumholz, M.~R. 2012, The Astrophysical Journal, 758, 48,
  \dodoi{10.1088/0004-637X/758/1/48}

\bibitem[{Yang {et~al.}(2018)Yang, Low, \& Johansen}]{Yang2018}
Yang, C.-C., Low, M.-M.~M., \& Johansen, A. 2018, The Astrophysical Journal,
  868, 27, \dodoi{10.3847/1538-4357/aae7d4}

\bibitem[{Youdin \& Johansen(2007)}]{Youdin_2007}
Youdin, A., \& Johansen, A. 2007, \apj, 662, 613, \dodoi{10.1086/516729}

\bibitem[{Youdin \& Goodman(2005)}]{Youdin2005}
Youdin, A.~N., \& Goodman, J. 2005, The Astrophysical Journal, 620, 459,
  \dodoi{10.1086/426895}

\bibitem[{Youdin \& Lithwick(2007)}]{YoudinLithwick2007}
Youdin, A.~N., \& Lithwick, Y. 2007, Icarus, 192, 588,
  \dodoi{10.1016/j.icarus.2007.07.012}

\bibitem[{{Youdin} \& {Shu}(2002)}]{Youdin_2002}
{Youdin}, A.~N., \& {Shu}, F.~H. 2002, \apj, 580, 494, \dodoi{10.1086/343109}

\bibitem[{{Zhuravlev}(2019)}]{Zhuravlev2019}
{Zhuravlev}, V.~V. 2019, \mnras, 2063, \dodoi{10.1093/mnras/stz2390}

\end{thebibliography}
\bibliographystyle{aasjournal}

\end{document}